\documentclass[10pt,english,review]{elsarticle}
\usepackage[T1]{fontenc}
\usepackage[latin9]{inputenc}
\usepackage[a4paper]{geometry}
\usepackage{array}
\usepackage{float}
\usepackage{units}
\usepackage{amsmath}
\usepackage{graphicx}

\makeatletter

\providecommand{\tabularnewline}{\\}

\usepackage{gensymb}
\usepackage{circuitikz}

\makeatother

\usepackage{babel}
\begin{document}

\title{GaN HEMT based high energy particles detection preamplifier}

\author[phys]{G. Orr\corref{corr}}
\ead{gilad.orr@ariel.ac.il}
\author[eng]{M. Azoulay}
\author[eng]{G. Golan}
\author[fisk]{A. Burger}
\author[]{}
\cortext[corr]{Corresponding author}
\address[phys]{Department of Physics, Ariel University, Ariel 40700, Israel}
\address[eng]{Department of Electrical Engineering, Ariel University, Ariel 40700, Israel}
\address[fisk]{Department of Physics, Fisk University, 1000, 17 Avenue North, Nashville, Tennessee 37208, USA}
\begin{abstract}
GaN high electron mobility transistors (HEMT) have gained some foothold
in the power electronics industry due to wide frequency bandwidth
and power handling. The material offers a wide bandgap and higher
critical field strength compared to most wide bandgap semiconductors,
resulting in better radiation resistance and theoretically higher
speeds as the devices dimensions could be reduced without suffering
voltage breakdown. This work consists of the underlying simulation
work intended to examine the response of the GaN HEMTs preamlifying
circuits for high resolution high energy radiation detectors. The
simulation and experimental results illustrate the superior performance
of the GaN HEMT in an amplifying circuit. Using a spice model for
a commercially available GaN HEMT non distorted output to an input
signal of 200 ps was displayed. Real world measurements underscore
the fast response of the GaN HEMT with its measured slew rate at approximately
$3000\,V/\mu s$ a result only $17\%$ lower than the result obtained
from the simulation. 
\end{abstract}
\maketitle

\section{Introduction}

GaN has been applied in power electronics as a replacement for Si-based
MOSFETs. As it has a wider band gap, higher mobility, i.e. lower resistance,
it provides lower power dissipation increasing its power handling
as compared to Si-based MOSFETs. Two additional advantages which have
not been investigated, are its higher radiation damage threshold and
its faster switching times. High radiation damage threshold increases
the components reliability in space and while incorporated into nuclear
detectors. When placing a GaN FET transistor at the input stage of
a radiation detector, both in addition to the larger band gap, can
result in an improved sensitivity and time resolution. Table \ref{tab:Electrical-and-physical}
displays a comparison of some important properties regarding Si, SiC
and GaN. 

\begin{table}[H]
\begin{centering}
\begin{tabular}{>{\raggedright}p{6.2cm}>{\centering}p{3.5cm}cccc}
\hline 
Property &  & Si & GaAs & SiC & GaN\tabularnewline
\hline 
Band gap & $E_{G}\,[eV]$\textsuperscript{\citep{wellmann2017power}} & 1.1 & 1.42 & 2.3\textasciitilde 3.3 & 3.44\tabularnewline
Critical field strength & $E_{C}\,[10^{6}V/cm]$\textsuperscript{\citep{wellmann2017power}} & 0.4 & 0.5 & 4 & 6\tabularnewline
Mobility & $\mu\,\left[\nicefrac{cm^{2}}{V\cdot s}\right]$\textsuperscript{\citep{microsemi2014gallium,ren2003wide}} & 1450 & 5000 & 900 & 2000\tabularnewline
Thermal conductivity & $\kappa\,\left[\nicefrac{W}{cm\cdot K}\right]$\textsuperscript{\citep{wellmann2017power}} & 1.5 & 0.5 & 3\textasciitilde 5 & 1.3\tabularnewline
Electron saturation velocity & $ve_{sat}\,\left[10^{7}\nicefrac{cm}{s}\right]$\textsuperscript{\citep{microsemi2014gallium,ren2003wide,ayzenshtat2011measurement}} & 1 & 1.4 & 2.2 & 3\tabularnewline
Lattice Constant  & $(\text{Å})$\textsuperscript{\citep{microsemi2014gallium,wellmann2017power}} & 5.43 & 5.65 & 3.08 & 3.19\tabularnewline
Coefficient of Thermal Expansion  & $\alpha\,[10^{-6}\times K^{-1}]$\textsuperscript{\citep{microsemi2014gallium,pierron1967CoefficientGaAs}} & 2.6 & 6.86 & 4.2 & 5.6\tabularnewline
\hline 
\end{tabular}
\par\end{centering}
\caption{\label{tab:Electrical-and-physical}Electrical and physical properties
of wide bandgap semiconductors }
\end{table}

The thermal conductivity $\kappa$ of the GaN is lower than Si and
SiC. The implication of it is that GaN has lower power dissipation,
limiting it to lower voltages, if one intends to use it for power
devices. Compensating for it is the high mobility $\mu$ which compounded
with a high charge (electron) carrier concentration, results in high
conductivity and low resistance. In the specific case of detectors,
the band gap $E_{G}$, critical field strength $E_{C}$ and mobility
$\mu$ are the dominating factors for an appropriate material while
the thermal conductivity $\kappa$ is less of a concern as not much
heat dissipation is expected. Overall, table \ref{tab:Electrical-and-physical}
indicates that excluding thermal conductivity, GaN properties promise
superior performance for high voltage fast switching of detectors. 

\section{GaN FET substrate}

While FET devices fabricated on Silicon, date back to the previous
century 50's, commercial SiC and GaN devices are relatively new comers,
with substantial improvements reducing the crystalline defects in
the materials. GaN substrates (free standing wafers) are currently
manufactured up to a diameter of approximately 50mm (2''), with a
company named Lumilog (Saint-Gobain Ceramics), recently offering 100
mm diameter wafers. Homoepitaxial growth enjoys lattice matching,
same thermal expansion coefficient between the substrate and epitaxial
layer, promoting improved crystal formation. Due to the excellent
structural properties homoepitaxy results in low dislocation densities
of $10^{3}-10^{5}\,cm^{-1}$ as compared heteroeptaxial growth, in
which threading dislocations with densities of $10^{9}-10^{10}$ when
grown on SiC or Sapphire substrates \citep{kamp1999gan}. Unfortunately,
growth of those freestanding substrates is challenging, there are
additional challenges associated with homoepitaxial GaN growth on
those substrates due to the chemical reactivity of the surfaces \citep{STORM2016121}.
This chemical reactivity results in their susceptibility to oxidation,
thermal decomposition and roughening. Those challenges in addition
to limited wafer diameter lead to the fact that prevalent substrates
for devices are hetroexpitaxy GaN structures grown on sapphire, SiC
and silicon. Of those substrates, SiC and silicon are being promoted
by leading manufacturers of GaN devices. In order to account for the
lattice mismatch, the use of buffer/matching layers is practiced.
$AlN$ and $AlGaN$ are common materials used for matching, though
being non native, still lead to stress and defects in the material.
The process begins by forming a GaN seed layer and a graded layer
of both AlGaN + GaN, over which GaN is grown \citep{kuzuhara2016algan}.
Another factor which may increase the defect density with time (thus
reducing reliability) is the difference in the thermal expansion (Table
\ref{tab:Electrical-and-physical}). While the thermal expansion coefficient
for GaN and SiC are at a ratio of 4:3 for GaN and Si they are 2:1
which is a significant drawback (and detrimental) on power devices
in which temperature varies periodically. GaN was examined and found
to be radiation resistant but prone to premature failure due to current
transients from the off state to the on state, resulting in an increase
in the dynamic $R_{on}$ leading to a current collapse exacerbated
at high temperatures \citep{golan2018novel,golan2018improved}. This
issue mitigated by surface passivation using AlN. We have to note
that due to the high mobility of the GaN, power FETs with low on resistance
reduce the component temperature increase at high currents due to
lower power dissipation. Given that the component does not act as
a power device and is not prone to temperature cycles, it's higher
immunity to radiation damage (radiation hardening) due to its wider
bandgap \citep{ren2003wide} is a virtue for detector electronics.
While devices using SiC as a substrate are commercially available,
Si substrate high electron mobility GaN transistors are by far a more
viable solution both due to available silicon microelectronics infrastructure
and experience.

\section{High Electron Mobility Transistor}

\begin{figure}[H]
\centering{}\includegraphics[scale=0.7]{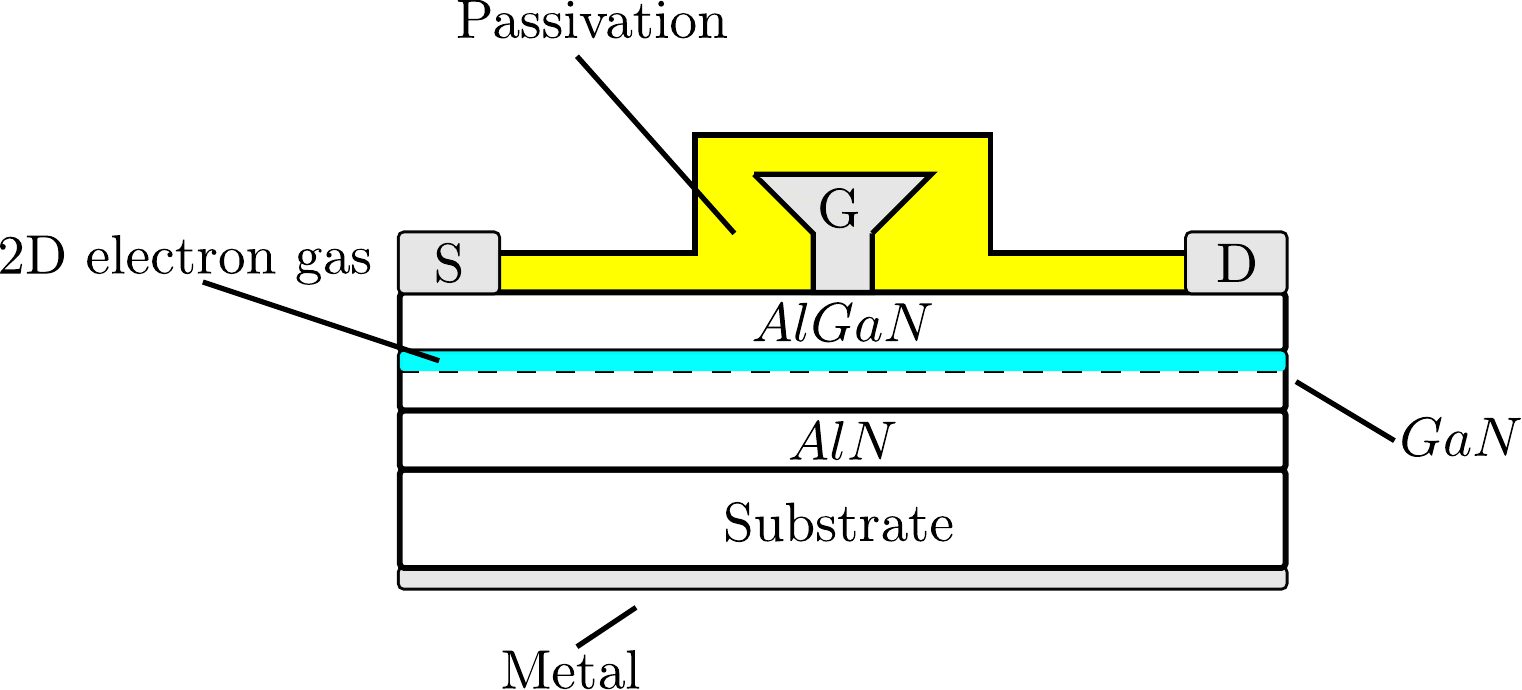}\caption{\label{fig:Basic-GaN/AlGaN-HEMT}Basic GaN/AlGaN HEMT}
\end{figure}

Similar to GaAs, GaN's frequency performance can be improved by adopting
a hetrostructure containing layers of AlGaN and GaN (similar to modulation
doping in GaAs) two layers of different semiconductors to which the
gate consists of a metal-semiconductor Schottky diode with the regular
contacts for drain and source (figure \ref{fig:Basic-GaN/AlGaN-HEMT})

Figure \ref{fig:Basic-GaN/AlGaN-HEMT} illustrates a basic GaN HEMT
transistor structure. The transistor hetrostructure is placed on a
substrate as discussed earlier in this article. An AlN buffer or nucleation
layer is placed on the substrate after which a GaN layer is acting
as the channel. An AlGaN barrier layer is deposited on top of the
GaN layer forming a hetrojunction. The hetrojunction offers an important
advantage as charge carriers are increased considerably without introducing
dopant impurities and without degrading the mobility. This structure
has the unique property in which charge carriers accumulate on the
asymmetric junction formed between the two semiconductors with different
band structures, confining them to the interface creating what is
known as a two dimensional electron gas (2DEG). Transistors consisting
of those structures are known as High Electron Mobility Transistors
(HEMT) with GaN transistors, theoretically showing promise of cutoff
frequencies up to the THz range. The important feature of nitrides
is that they have a wurtzite crystal structure, which has a lower
symmetry and hence built in polarization. In adition to this, when
growing AlGaN on a GaN layer the AlGaN is stressed, creating piezoelectric
polarization as it has a wider lattice structure than GaN. This adds
to the spontaneous polarization built into the unstrained crystal
resulting in a sheet of uncompensated positive charge at the interface
creating a quantum well at the interface, leading to carrier confinement
and a 2D electron gas in the undoped material as well as the doped
one. This is in contrast to doped semiconductors such as GaAs and
InP in which the doping itself is responsible for ionized impurity
scattering, reducing mobility. The 2D electron gas in GaN has a sheet
charge density of the order of $10^{13}\,cm^{-2}$ that is five times
larger than a similar device using GaAs/AlGaAs. A detailed account
including the band structure can be found in \citep{roccaforte2019overview}.
There are other challenges that are still being ironed out with the
GaN HEMT, resulting for example from surface stages on the AlGaN and
traps near the interface in the GaN layer. Those things affect the
mobility or create a virtual gate and are addressed structurally in
various ways. For example, the passivation layer is intended to resolve
the AlGaN surface states. There is an ongoing research focused on
improvement of the crystalline layer purity and fabrication processes
for reducing the density of the AlGaN buffer traps. 

At this point, it is the time to place a distinction between small
signal HEMTs which are important for this current discussion to HEMTs
used in high frequency power switching. Current small signal devices
begin at frequencies above $40\,GHz$, while most of the available
work discusses power devices with larger gate lengths ($>0.5\mu$
) and larger device cross sections with lower cutoff frequencies.
The higher breakdown voltage of the GaN (table \ref{tab:Electrical-and-physical})
contributes to even smaller devices and gate sizes improving performance
at high frequencies.

\subsection{Operating frequency}

High capacitance and low impedance are directly correlated to the
transistors operating frequency. Increasing the cross section of the
device, results in an increase in its capacitance as the capacitance
is proportional to the area of the conductors, and at the same time
reduces the impedance for the current increases ($z=v/i$). Thus for
our purposes two frequencies characterize the device the gain bandwidth
product (GBP) the maximum frequency of oscillation $f_{max}$
\begin{align}
GBP & =\frac{g_{m}}{2\pi C_{gs}}\label{eq:GBP}\\
f_{max} & =\frac{GBP}{2}\sqrt{\frac{R_{ds}}{R_{g}}}\label{eq:f_max}
\end{align}
where $g_{m}$ is the transconductance, $C_{gs}$ the capacitive coupling
between the gate and channel, $R_{ds}$ a resistance modeling the
drain source current, and $R_{g}$ the gate resistance. The GBP is
the frequency in which is the short circuit current gain falls to
unity while $f_{max}$ is the frequency in which the power gain falls
to unity. The higher those values are the higher performance the device
provides at higher frequencies. If the area of the device is larger
it would increase the gate source capacitance $C_{gs}$ reducing the
GBP (equation \ref{eq:GBP}). This in turn will affect the maximum
frequency (equation \ref{eq:f_max}), which would reduce $R_{ds}$
that depends on the device area as well. Thus reducing the dimensions
inherently increases the operating frequencies. One property of the
material that affects the frequencies is the dielectric constant,
as 
\[
C=\varepsilon_{0}\varepsilon_{r}\frac{A}{d}
\]
a lower dielectric constant results in lower capacitive loading, thus
increasing the operating frequency. Table \ref{tab:Dielectric-constant-of}
compares the dielectric constant of several semiconducting materials

\begin{table}[H]
\centering{}%
\begin{tabular}{cc}
\hline 
Material & $\varepsilon_{r}$\textsuperscript{\citep{pengelly2012review}}\tabularnewline
\hline 
GaAs & 12.5\tabularnewline
InP & 12.4\tabularnewline
Si & 11.9\tabularnewline
SiC & 10.0\tabularnewline
GaN & 9.5\tabularnewline
Diamond & 5.5\tabularnewline
\hline 
\end{tabular}\caption{\label{tab:Dielectric-constant-of}Dielectric constant of several
common semiconducting materials}
\end{table}
As can be seen, in this sense, GaN is favorable to other semiconducting
materials excluding diamond. One has to recall that once we discuss
the GaN HEMTs, the gate consists of a Shockley diode further reducing
the gate channel capacitive coupling. Returning to table \ref{tab:Electrical-and-physical},
it can be seen that both electron mobility and electron saturation
velocity are higher in GaN than in SiC and Si, positioning it as the
better material for high frequency solid state devices. To this we
can add the polarized HEMT structure which is currently commercially
unique to GaN providing much higher mobility compared to other materials
including GaAs with a similar structure. It has a wider bandgap compared
to GaAs which improves its reliability in environments with energetic
radiation and a considerably lower noise figure \citep{harris2011commercial},
placing it in an attractive replacement in the first stage of the
preamplifier. Here we have to state again, that we are interested
that the device will obtain the higher frequency and not high frequency
power devices for which the SiC may in certain circumstances be the
better material due to its superior thermal conductivity. 

In order to get some idea of the frequencies involved, some recent
work should be referenced. Kabouche et. al. have reported an AlN/GaN
HEMT with a maximum frequency $f_{max}=242\,GHz$ \citep{kabouche2019high},
with Li et. al reporting a cutoff frequency, GBP of $204\,GHz$ and
maximum device frequency of $f_{max}=250\,GHz$ \citep{li2020gan}.
This follows the work done by Y. Durmus obtaining GBP of $100\,GHz$
and $f_{max}=128\,GHz$ using AlGaN/GaN. While this displays frequencies
nearing the THz range, available commercial devices which are designed
mostly for power amplification, are available for cutoff frequencies
of up to 20 GHz such as work from Kim et. al. with devices attaining
GBP of approximately $19\,GHz$ and $f_{max}$ of $60\,GHz$ \citep{kim2020effects}.

\section{GaN HEMT and silicon JFET detection circuits (Preamplifiers)}

\begin{figure}[H]
\centering{}%This is the drawing TeX file of the JFET and BJT circuit

\begin{circuitikz}
    \fontsize{10}{10}
    \draw (0,0) to [battery1, l=$HV$] (0,3);
    \draw (3,0) to [american resistor, a=$\begin{array}{l}R_1\\20M\Omega\end{array}$, name = R1] (3,3);
    \draw (6,0) to [american resistor, a=$\begin{array}{l}R_2\\2K\Omega\end{array}$] (6,2.5);
    \draw (6,3.28) node [njfet, ]{};
    \draw (9,2.5)  node[npn]{};
    \draw (6,2.5) to (8.5,2.5);
    \draw (9,0) to [american resistor, a=$\begin{array}{l}R_3\\200\Omega\end{array}$] (9,2);
    \draw (9,3) to [american resistor, a=$\begin{array}{l}R_4\\2K\Omega\end{array}$] (9,6);
    \draw (0,3) to [PZ, l=$Detector$] (0.75,3) to (5.2,3);
    \draw (6,4) to (6,6) to (9,6);
    \draw (0,0) to (12.3,0); 
    \draw(12.3,0) to (12.3,3.22);
    \draw(9,3.5) to [capacitor, a_=$100pF$,l^=$C_1$] (12,3.5) node[bnc, scale=2, rotate=180]{};
    %\draw(12,3.5) -- (13,3.5) node[ocirc]{} ;
    %\draw(12,0) -- (13,0) node[ocirc]{} ;
    \draw (6,0) node [ground]{};
    \draw (7.5,6) node[vcc, ]{+12V};
    %Q1 text
    \node [align=left] at (6.7, 3.25) {J1\\2N4416};
    %Q2 text
    \node [align=left] at (9.7,2.5) {Q1\\2N3904};
\end{circuitikz}\caption{\label{fig:Simple-detection-circuit}Simple preamplifier using a JFET
transistor in its first stage}
\end{figure}
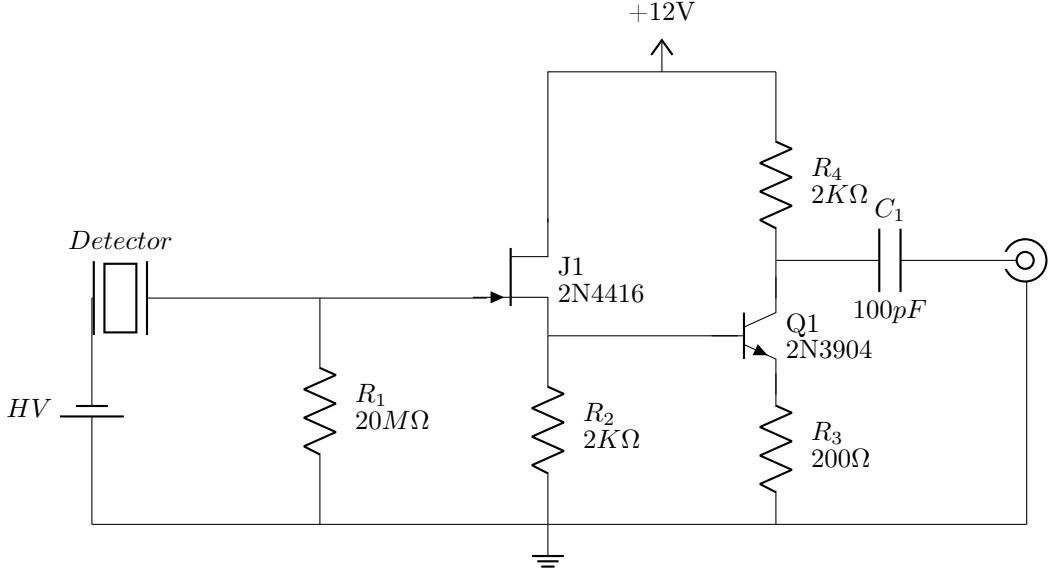

The conventional way of sensing the $\gamma$-ray energy consists
of using a JFET transistor (2N4416) as a source follower at the first
stage tracing the voltage of the charge generated by the interaction
with the $\gamma$ photon and the detector. JFET's are used due to
their speed, low noise and high input impedance. A simple detection
circuit, based on \citep{zuck2004microstructure} is illustrated in
figure \ref{fig:Simple-detection-circuit}. 
In the above circuit, the HV bias controls the sensitivity of the
circuit at times requiring the use of a second amplification stage
consisting of the 2N3904 transistor. While the first stage transistor
has a switching speed of $2.5\,ns$, the actual measured time resolution
at its output is approximately $20\,ns$. The switching speed of the
second general small signal amplifying transistor, is approximately
$250\,ns$ providing a larger signal (small signal amplification)
at the expense of the time resolution. This is followed by an amplifier
with the provisions of shaping the signal obtained from the pre-amplifier.
For pulse height spectroscopy an A/D and a multichannel analyzer are
connected to the output. As we have seen earlier in this article,
GaN based transistors are faster, thus replacing both stages with
GaN transistors, one can expect that the time resolution would improve
considerably. Having a wide bandgap GaN based transistors tend to
be radiation hardened compared to the above circuit, increasing its
reliability in environments where such radiation is abundant, such
as space. Placing a GaN based FET at the first preamplification stage
is attractive as it may not require the radiation shielding associated
with shallow bandgap semiconductors, thus resulting in a reliable
and compact circuit. 

\subsection{Preamplifier simulations}

Prior to realizing a GaN circuit, its merits should be analyzed and
different circuits including a GaN model need to be compared. The
simulation was carried using LTspice XVII with a spice model of the
fastest GaN HEMT that Efficient Power Conversion offers commercially
\citep{EPC2038}. Three circuits were compared, a preamplifier with
a JFET input and bipolar transistor at its output (Figure \ref{fig:Simple-detection-circuit}),
a preamplifier with a JFET at its input and an NMOS at its output
and a circuit with an eGaN. The following simulations are for comparing
theoretical features only. This will be followed by the measured output
of practical circuits compared with the simulated result based on
the measurement systems constraints. One has to remember, that we
are testing the response of the system to a short pulse and in the
following instances the simulated input pulses are large, driving
the circuit into saturation, rendering the second stage as redundant,
seemingly reducing the voltage amplification. In practice most of
the pulses will have a much lower amplitude requiring the additional
amplification. There will be instances in which using only the first
stage will have its benefits especially when a response to very short
pulses are a concern.

\subsubsection{JFET input and bipolar transistor at the output}

\begin{figure}[H]
\begin{centering}
%This is the drawing TeX file of the JFET and BJT circuit
\begin{circuitikz}
    \fontsize{10}{10}
    \draw (0,0) to [sqV, l=$V_{pulse}$, a=$\begin{array}{l}V_{+}{=}10V\\V_{-}{=}{-}10V\\t_{+}{=}3ns\\f{=}5MHz\end{array}$] (0,3);
    \draw (3,0) to [american resistor, a=$\begin{array}{l}R_1\\20M\Omega\end{array}$, name = R1] (3,3);
    \draw (6,0) to [american resistor, a=$\begin{array}{l}R_2\\2K\Omega\end{array}$] (6,2.5);
    \draw (6,3.28) node [njfet, ]{};
    \draw (9,2.5)  node[npn]{};
    \draw (6,2.5) to (8.5,2.5);
    \draw (9,0) to [american resistor, a=$\begin{array}{l}R_3\\200\Omega\end{array}$] (9,2);
    \draw (9,3) to [american resistor, a=$\begin{array}{l}R_4\\2K\Omega\end{array}$] (9,6);
    \draw (0,3) to (5.2,3);
    \draw (6,4) to (6,6) to (9,6);
    \draw (0,0) to (12,0); 
    \draw(12,0) to [american resistor, l_=$\begin{array}{l}R_5\\1M\Omega\end{array}$] (12,3.5);
    \draw(9,3.5) to [capacitor, a_=$100pF$,l^=$C_1$] (12,3.5);
    \draw(12,3.5) -- (13,3.5) node[ocirc]{} ;
    \draw(12,0) -- (13,0) node[ocirc]{} ;
    \draw (6,0) node [ground]{};
    \draw (7.5,6) node[vcc, ]{+12V};
    %Q1 text
    \node [align=left] at (6.7, 3.25) {J1\\2N4416};
    %Q2 text
    \node [align=left] at (9.7,2.5) {Q1\\2N3904};
\end{circuitikz}
\par\end{centering}
\caption{\label{fig:A-preamplifier-detection}A preamplifier detection simulation
circuit with a JFET at the first stage and a generic bipolar transistor
at the second stage.}
\end{figure}
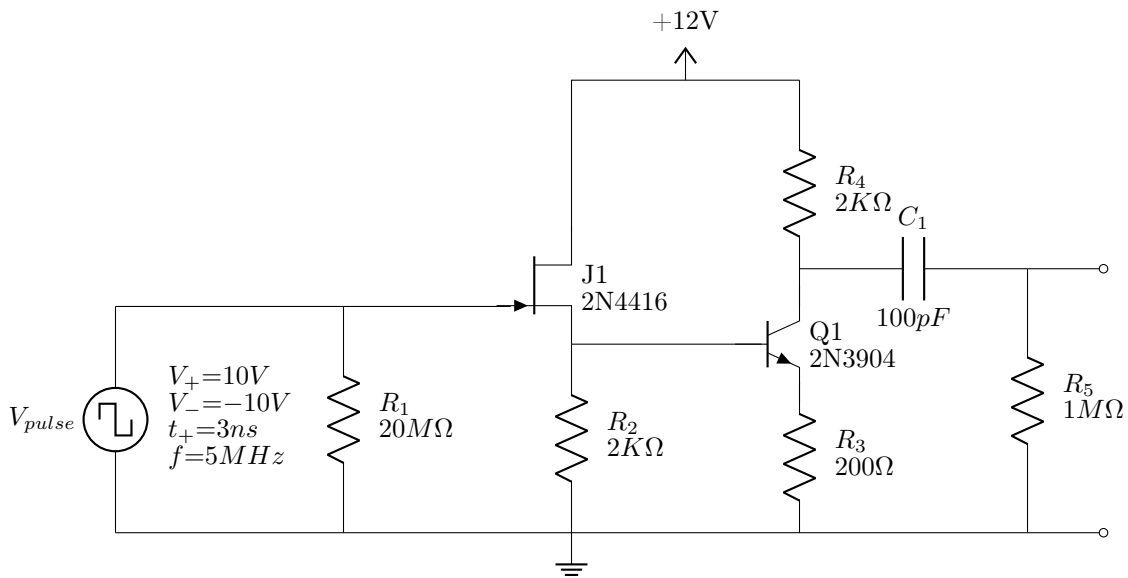

Figure \ref{fig:A-preamplifier-detection} displays the circuit simulated
of a preamplifier circuit with a JFET at the input (2N4416) and a
generic bipolar transistor at the output (2N3904). The input stage
consists of a high impedance source follower further amplified by
the generic transistor.

\begin{figure}[H]
\begin{centering}
\includegraphics{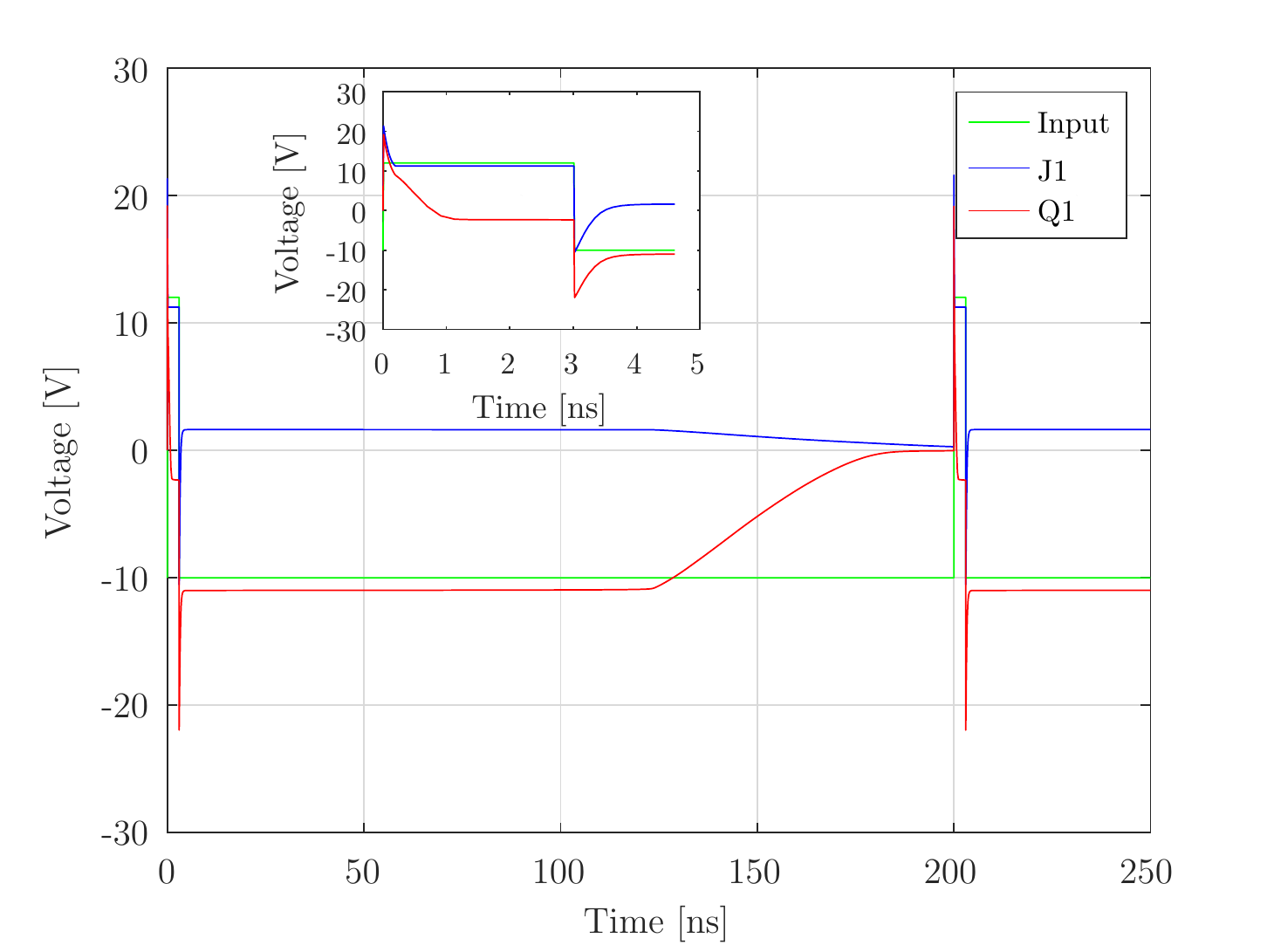}
\par\end{centering}
\caption{\label{fig:Simulation-of-the-JFET-Bipolar}Simulation of the response
of the circuit with a JFET at its input and a generic transistor at
its output}
\end{figure}

Instead of the detector we inserted a pulse generator with a pulse
height of 20V, pulse width of 3ns and a period of 200ns. Figure \ref{fig:Simulation-of-the-JFET-Bipolar}
displays the simulation results of the pulse at the output of the
JFET (J1) and the generic transistor (Q1).

From the simulation it is obvious that the reverse recovery charge
$Q_{rr}$ of the JFET and the resulting current $I_{rr}$, lead to
spikes in the drain-source voltage. It is obvious as well that while
the JFET's response tracks the pulse fairly well, a residual voltage
at the output results in Q1 conducting increasing the response time
of the circuit to approximately $200\,ns$. In the inset the response
of both transistor outputs at the duration of the pulse are shown. 

\subsubsection{JFET input and NMOS output}

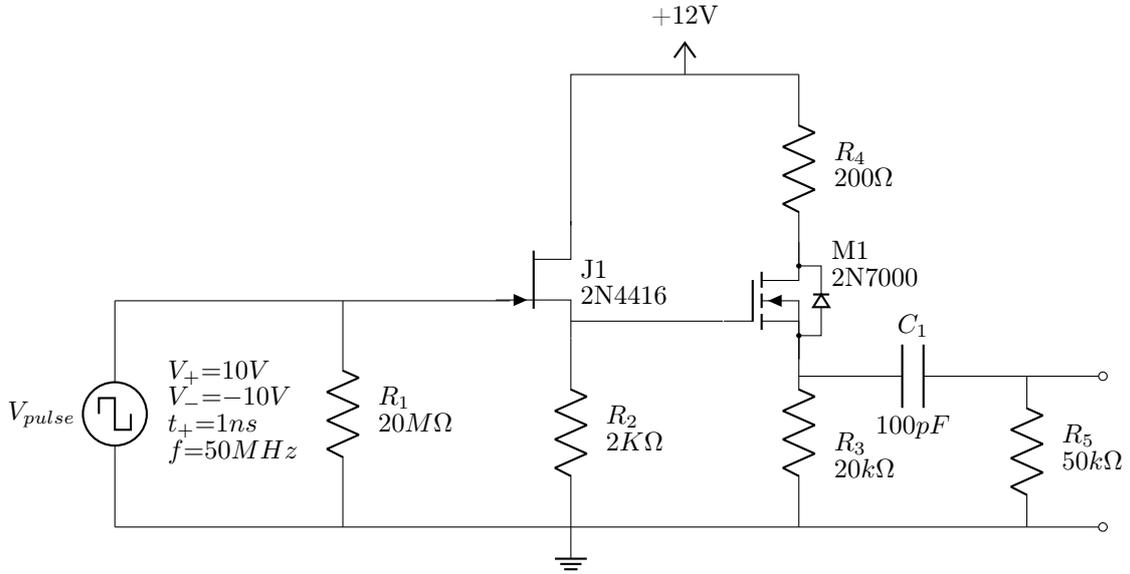
\begin{figure}[H]
\begin{centering}
%This is the drawing TeX file of the JFET and NMOS circuit
\begin{circuitikz}
    \fontsize{10}{10}
    \draw (0,0) to [sqV, l=$V_{pulse}$, a=$\begin{array}{l}V_{+}{=}10V\\V_{-}{=}{-}10V\\t_{+}{=}1ns\\f{=}50MHz\end{array}$] (0,3);
    \draw (3,0) to [R, a=$\begin{array}{l}R_1\\20M\Omega\end{array}$, name = R1] (3,3);
    \draw (6,0) to [R, a=$\begin{array}{l}R_2\\2K\Omega\end{array}$] (6,2.5);
    \draw (6,3.28) node [njfet, ]{};
    \draw (9,3)  node[nigfete, bodydiode]{};
    \draw (6,2.73) to (8,2.73);
    \draw (9,0) to [R, a=$\begin{array}{l} \\ \\R_3\\20k\Omega\end{array}$] (9,2.5);
    \draw (9,3.5) to [R, a=$\begin{array}{l}R_4\\200\Omega\end{array}$] (9,6);
    \draw (0,3) to (5.2,3);
    \draw (6,4) to (6,6) to (9,6);
    \draw (0,0) to (12,0); 
    \draw(12,0) to [R, l_=$\begin{array}{l}R_5\\50k\Omega\end{array}$] (12,2);
    \draw(9,2) to [capacitor, a_=$100pF$,l^=$C_1$] (12,2);
    \draw(12,2) -- (13,2) node[ocirc]{} ;
    \draw(12,0) -- (13,0) node[ocirc]{} ;
    \draw (6,0) node [ground]{};
    \draw (7.5,6) node[vcc, ]{+12V};
    %Q1 text
    \node [align=left] at (6.7, 3.25) {J1\\2N4416};
    %Q2 text
    \node [align=left] at (10,3.5) {M1\\2N7000};
\end{circuitikz}
\par\end{centering}
\caption{\label{fig:A-preamplifier-detection-JFET-NMOS}A preamplifier detection
simulation circuit with a JFET at the first stage and a generic enhanced
NMOS FET at the output stage}
\end{figure}

At this stage we wanted to test whether an improvement may be achieved
using an NMOS at the ouput. Figure \ref{fig:A-preamplifier-detection-JFET-NMOS}
illustrates the circuit used to simulate the preamplifier with the
same JFET at the input but with a generic enhanced NMOS transistor
at the output (2N7000). As the output consists of a source follower
we do not expect a shift in the output.

\begin{figure}[H]
\begin{centering}
\includegraphics{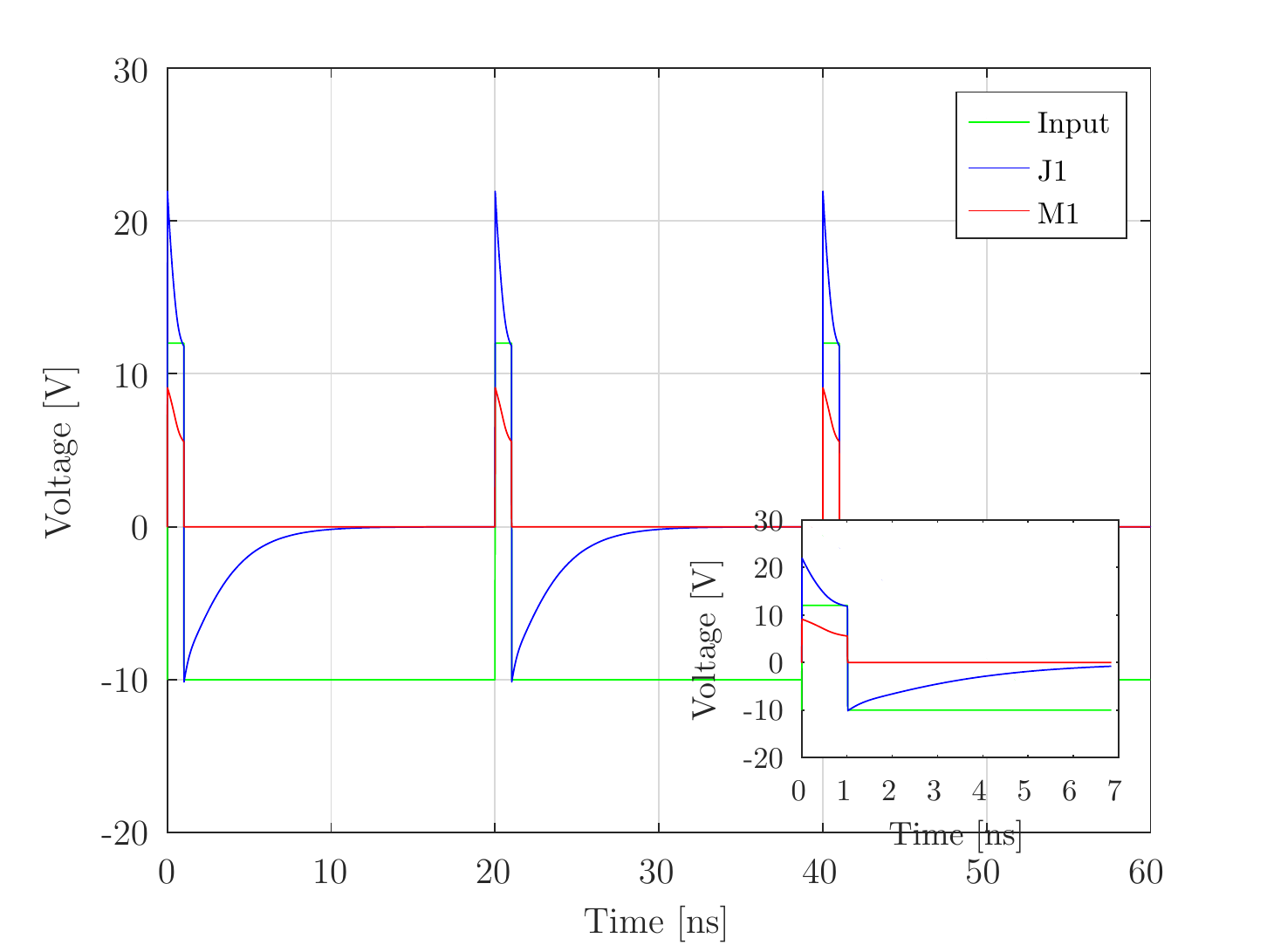}
\par\end{centering}
\caption{\label{fig:Simulation-of-the-response-JFET_input_NMOS_output}Simulation
of the response of the circuit with a JFET at its input and NMOS FET
at its output}
\end{figure}

Once again, the circuit's response was tested with a periodic short
pulse, but as this circuit display a faster response the pulse width
was reduced to $1\,ns$ with a period of $20\,ns$. Figure \ref{fig:Simulation-of-the-response-JFET_input_NMOS_output}
displays the resulting signals at the output of the JFET (J1) and
the NMOS FET (M1).

In this instance it is seen that the small signal NMOS FET does not
display the negative voltage induced reverse recovery charge behavior
while maintaining the width of the input pulse, though it does not
completely track its form. It can be seen from Figure \ref{fig:Simulation-of-the-response-JFET_input_NMOS_output}
that this has mostly to do with the JFET at the input.

\subsubsection{GaN HEMT preamplifier}

\begin{figure}[H]
\begin{centering}
%GaN child
\begin{circuitikz}
    \draw (0,0) to [sqV, l=$V_{pulse}$, a=$\begin{array}{l}V_{+}{=}10V\\V_{-}{=}{-}10V\\t_{+}{=}200ps\\f{=}666MHz\end{array}$] (0,3.5);
    \draw (3,0) to [R, a=$\begin{array}{l}R_1\\20M\Omega\end{array}$] (3,3.5);
    \draw (0,3.5) to (5.3,3.5);
    \draw (6,3.5) node [hemt]{};
    \draw (6,0) to [R, a=$\begin{array}{l}R_2\\10K\Omega\end{array}$] (6,2.5);
    \draw (6,2.5) to [C, l=$C_1$, a=$100p$] (9,2.5);
    \draw (9,0) to [R, a=$\begin{array}{l}R_3\\1M\Omega\end{array}$] (9,2.5);
    \draw (9,2.5) to (10,2.5) node[ocirc]{};
    \draw (0,0) to (10,0) node[ocirc]{};
    \draw (4.5,0) node [ground]{};
    \draw (6,4) to (6,5) node [vcc] {+12V};
    \draw (6,2.5) to (6,3);
    %Q1 text
    \node [align=left] at (7.1, 4) {Q1\\EPC2038};
\end{circuitikz}
\par\end{centering}
\centering{}\caption{\label{fig:A-GaN-HEMT}A GaN HEMT based two stage preamplifier with
a GaN HEMT transistor at the input and an NMOS FET at the output}
\end{figure}
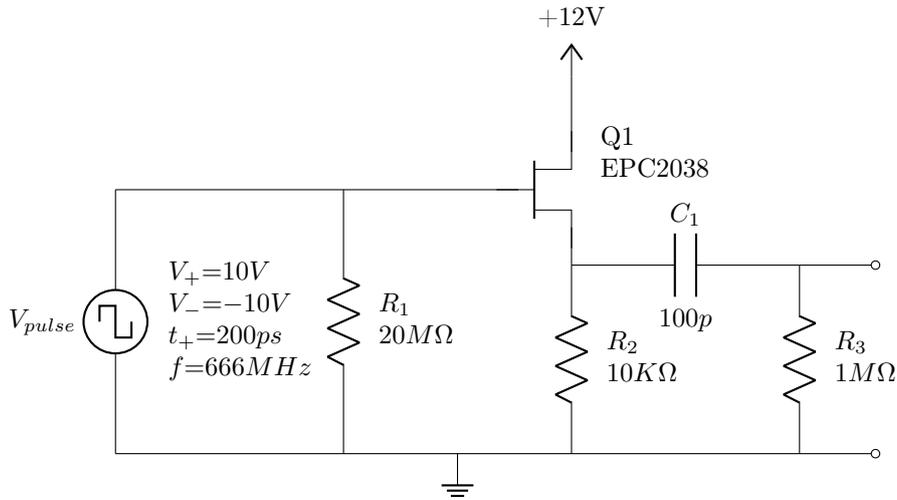

Figure \ref{fig:A-GaN-HEMT} displays a preamplifier circuit based
on a GaN transistor (EPC2038). The transistor is wired as a source
follower.

\begin{figure}[H]
\begin{centering}
\includegraphics[scale=0.7]{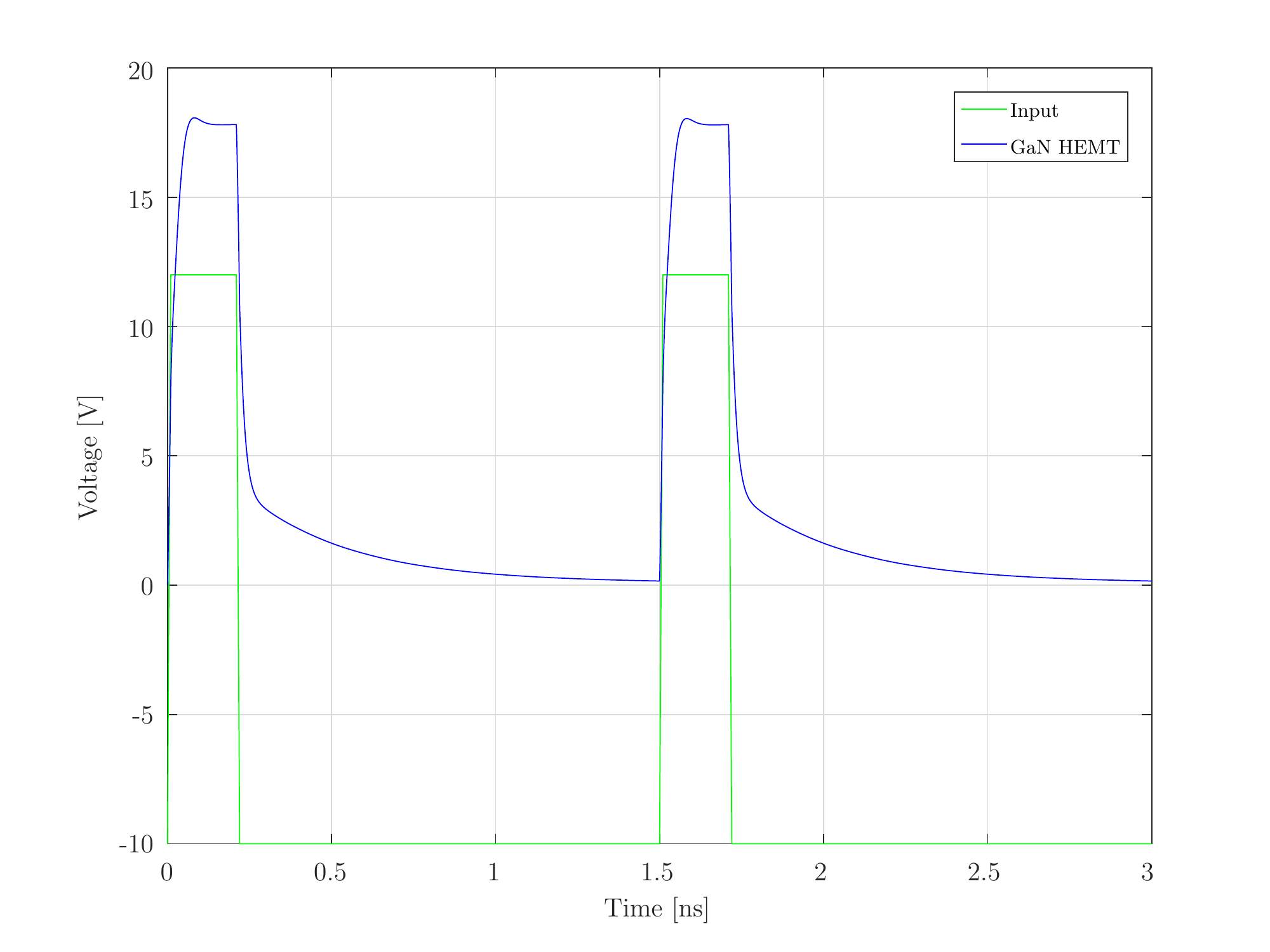}
\par\end{centering}
\caption{\label{fig:Response-simulation-of-GaN-NMOS}Response simulation of
the circuit with a GaN HEMT at its input and NMOS FET at its output}
\end{figure}

This circuit is expected to display a faster response. The test signal
consists of a 200ps wide pulse with a 1.5 ns period. Figure \ref{fig:Response-simulation-of-GaN-NMOS}
displays the resulting signals at the output of the GaN HEMT (U1)
and the NMOS FET (M1).

The simulation of this configuration shows a much better response
tracking the input signal with an acceptable fidelity at the output. 

\subsection{Practical amplifier circuits}

Once we have some understanding of the preamplifier circuit feasibility
and expected response, we proceeded to realizing them as physical
circuits. The circuits were an exact implementation of the simulated
circuits excluding the GaN HEMT in which the second stage was omitted.
An Agilent 33250 80MHz Function/Arbitrary waveform generator which
can go all the way down to a $8ns$ pulse, was used as a test input
signal for the various amplifiers. It was set to a $5MHz$ repetition
rate with $8,\,10,12,15,20,25,30,40$ and $60\,ns$ pulse widths.
The pulses amplitude was $20V$ rising from -10V to 10V. The generator
features a minimum pulse width of 8ns and a minimum rise time of 5ns.
Sampling was conducted using a Rohde Schartz RTM3004, 10Bit, 5GSa/s
with a 1 GHz bandwidth. RT-ZP05, 10:1 ratio probes were used having
a 500MHz bandwidth, impedance of $10M\Omega/10pF$ and a typical rise
time of 0.7 ns. Four points were sampled, at the input, at the output
of the first amplifying stage, at the output of the second amplifying
stage (with the DC bias) and at the preamplifiers output beyond the
$100pF$ output capacitor. The following graphs for the various circuits
illustrate their response to the generators pulse, given its minimum
pulse width and rise time. The response of the physical circuit is
compared to the simulated response cross validating both the circuits
and the simulation.

\subsubsection{JFET input with a bipolar transistor at the output}

\begin{figure}[h]
\begin{centering}
\includegraphics[scale=0.7]{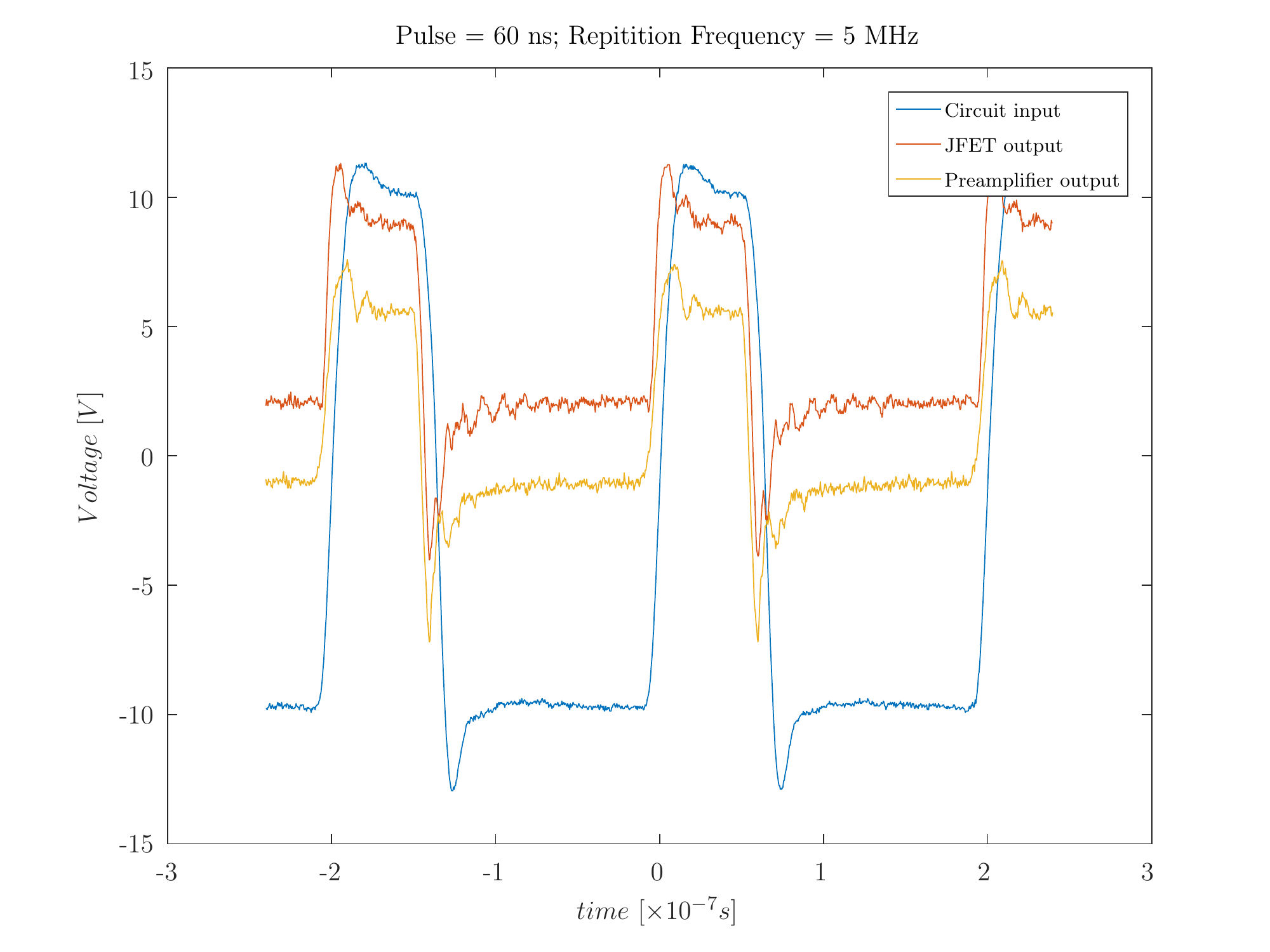}
\par\end{centering}
\centering{}\includegraphics[scale=0.7]{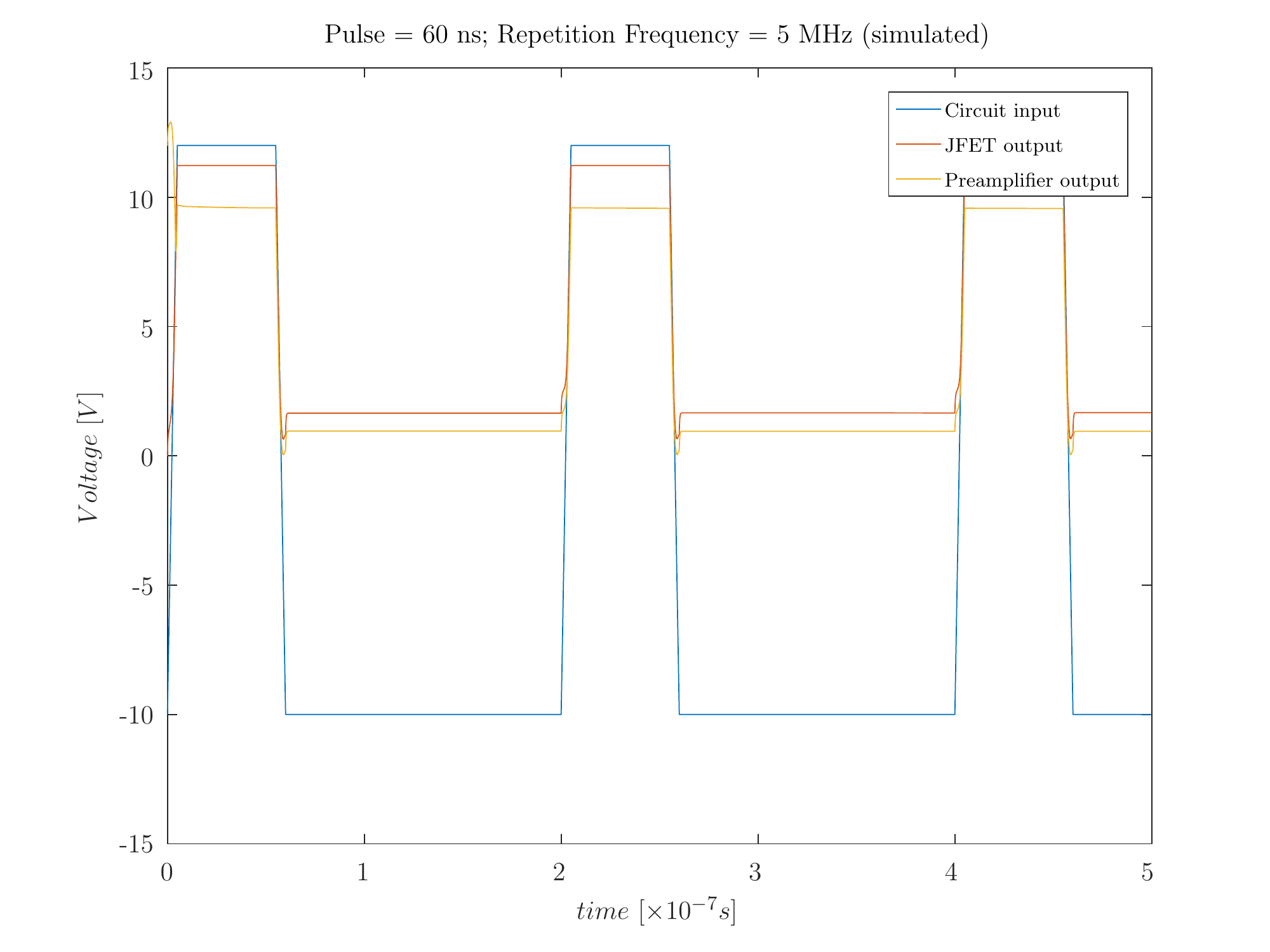}\caption{\label{fig:JFET-Bipolar-transistor-preampli}JFET-Bipolar transistor
preamplifier response to $60ns$ pulses}
\end{figure}

\begin{figure}[h]
\begin{centering}
\includegraphics[scale=0.7]{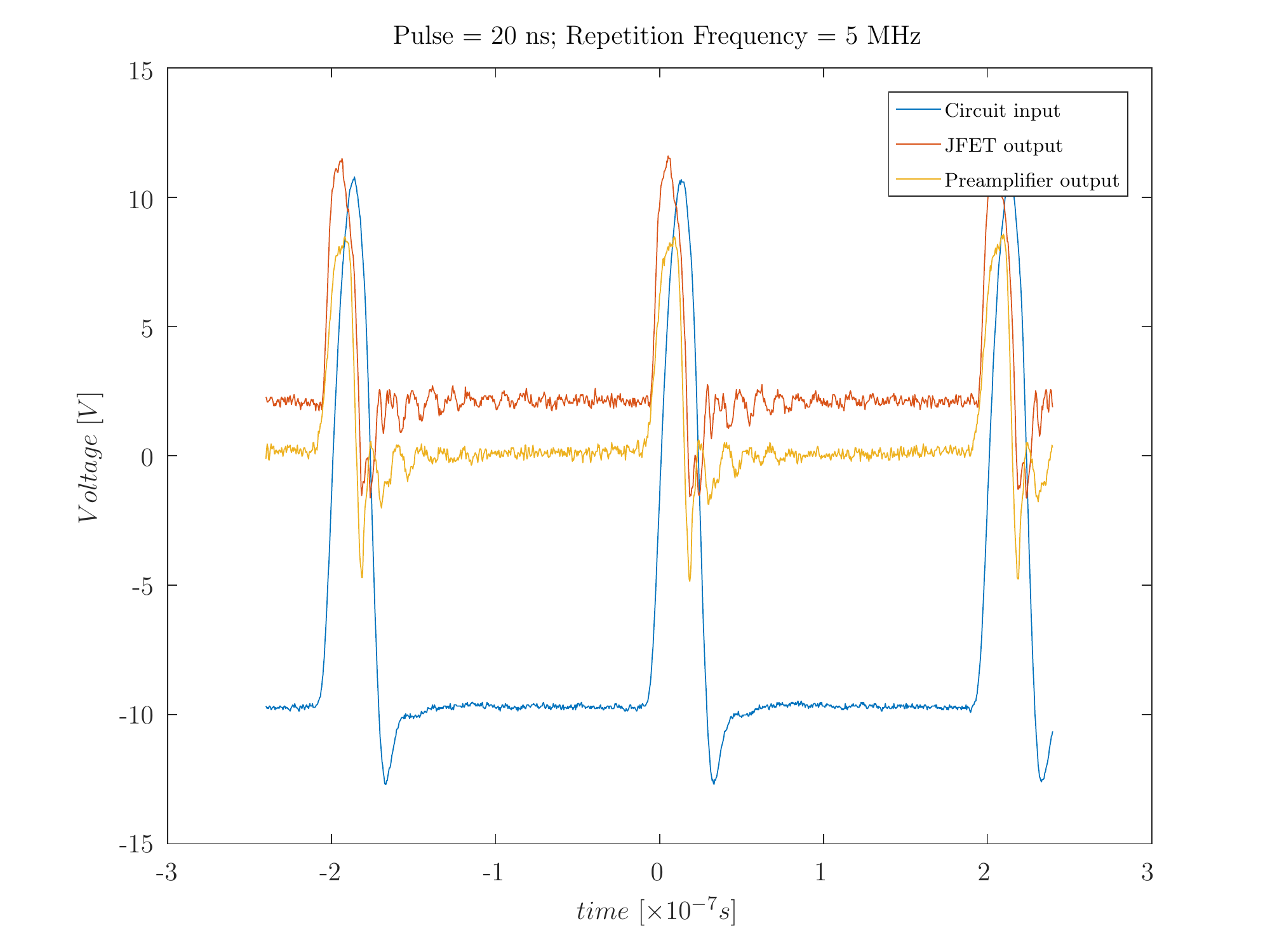}
\par\end{centering}
\centering{}\includegraphics[scale=0.7]{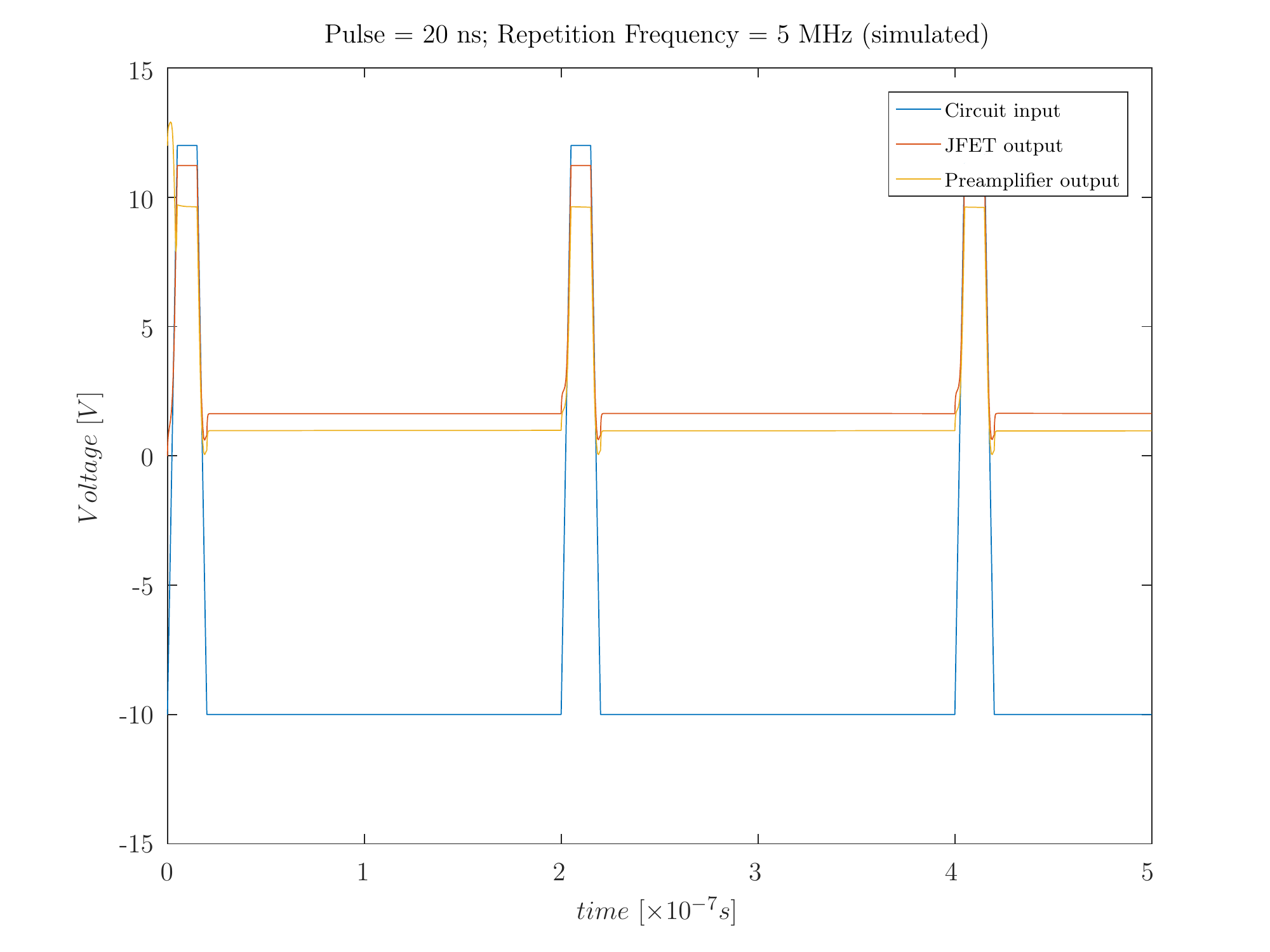}\caption{\label{fig:JFET-Bipolar-transistor-preampli-1}JFET-Bipolar transistor
preamplifier response to $20ns$ pulses}
\end{figure}

\begin{figure}[h]
\begin{centering}
\includegraphics[scale=0.7]{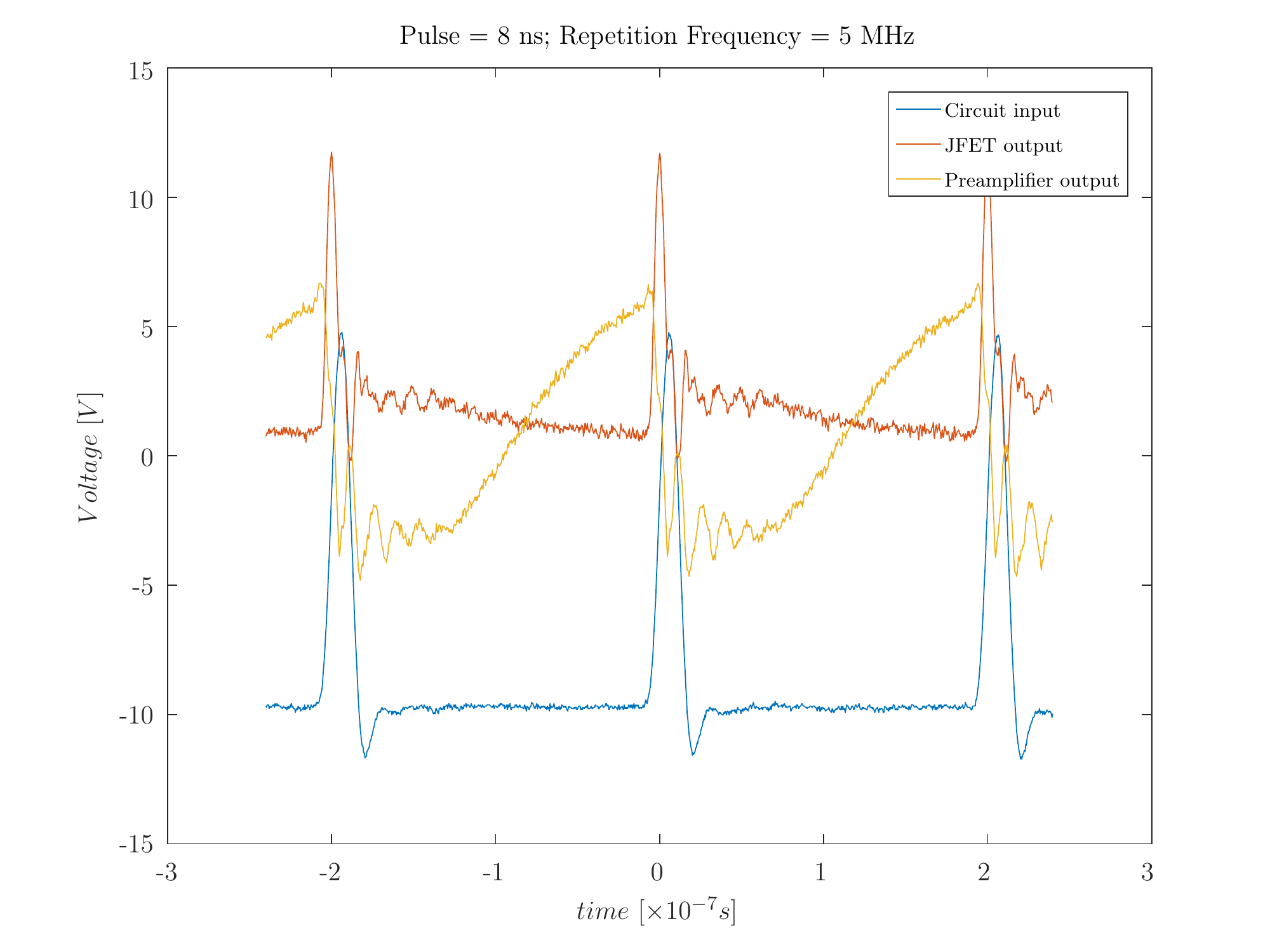}
\par\end{centering}
\begin{centering}
\includegraphics[scale=0.7]{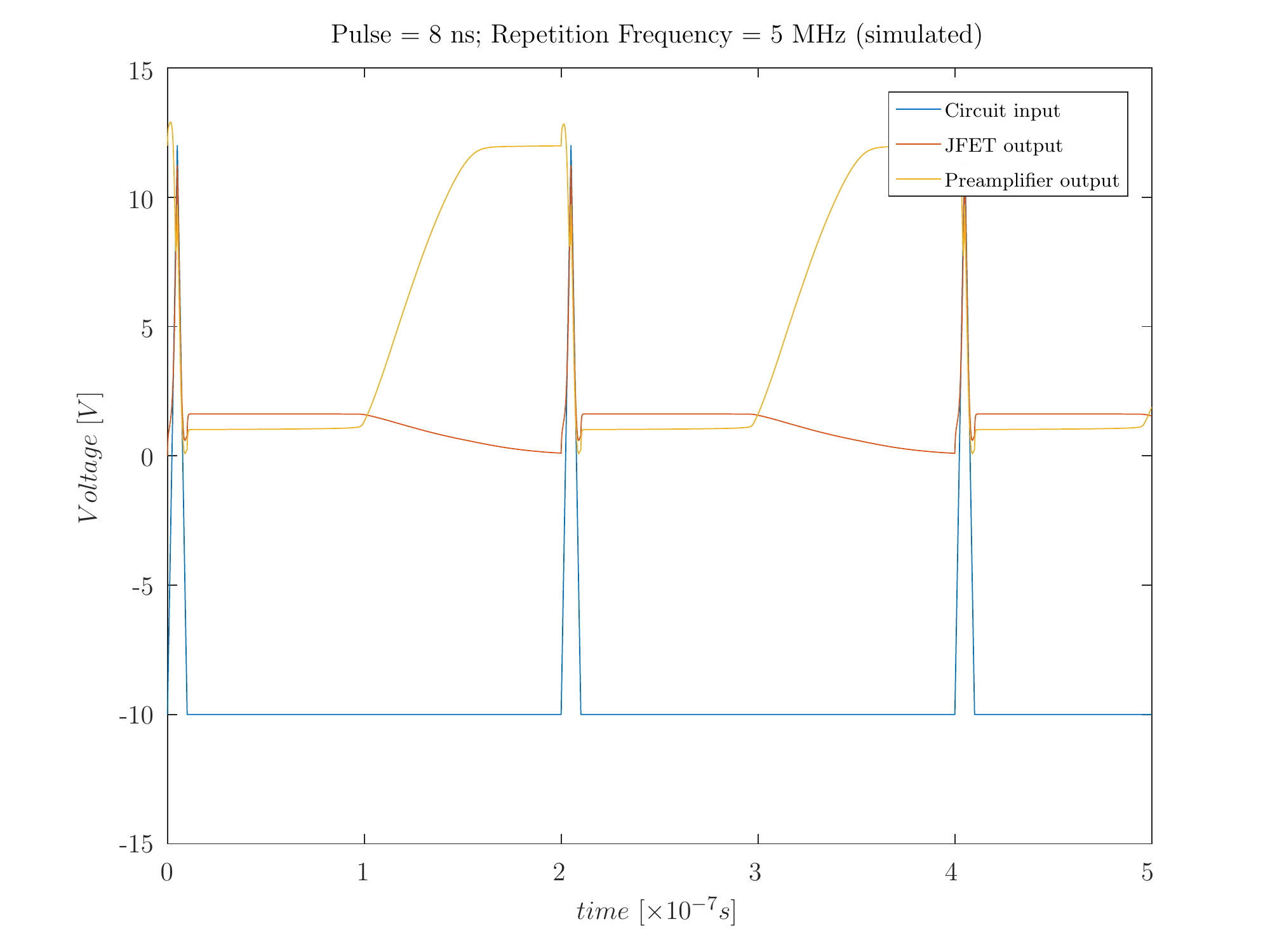}
\par\end{centering}
\centering{}\caption{\label{fig:JFET-Bipolar-transistor-preampli-2}JFET-Bipolar transistor
preamplifier response to $8ns$ pulses}
\end{figure}

We begin by demonstrating measurements regarding the circuit with
the JFET at its input stage (Figure \ref{fig:A-preamplifier-detection}).
At the second stage we used a standard general 2N3904 NPN transistor.
At the input stage we used a mil-spec 2N4416 N-Channel JFET in a TO206AF
hermetically-sealed package. This is an important feature which will
be discussed during stress testing in a following article. Figures
(\ref{fig:JFET-Bipolar-transistor-preampli}-\ref{fig:JFET-Bipolar-transistor-preampli-2})
illustrate the measured response of the preamplifier to a repetitive
$5MHz$ pattern pulses of $60,24$ and $8\,ns$ given the generator
limitations discussed above.

In all of the circuits the fast rise time and tracking of the JFET
at the first stage is evident. Compared to the $5\,ns$ rise time
of the pulse generator its response is almost instantaneous. This
does not apply to the general type transistor which is evident from
the response to the $8\,ns$ input in both the physical implementation
of the circuit and the simulation.

\subsubsection{JFET input and NMOS output}

\begin{figure}[h]
\begin{centering}
\includegraphics[scale=0.7]{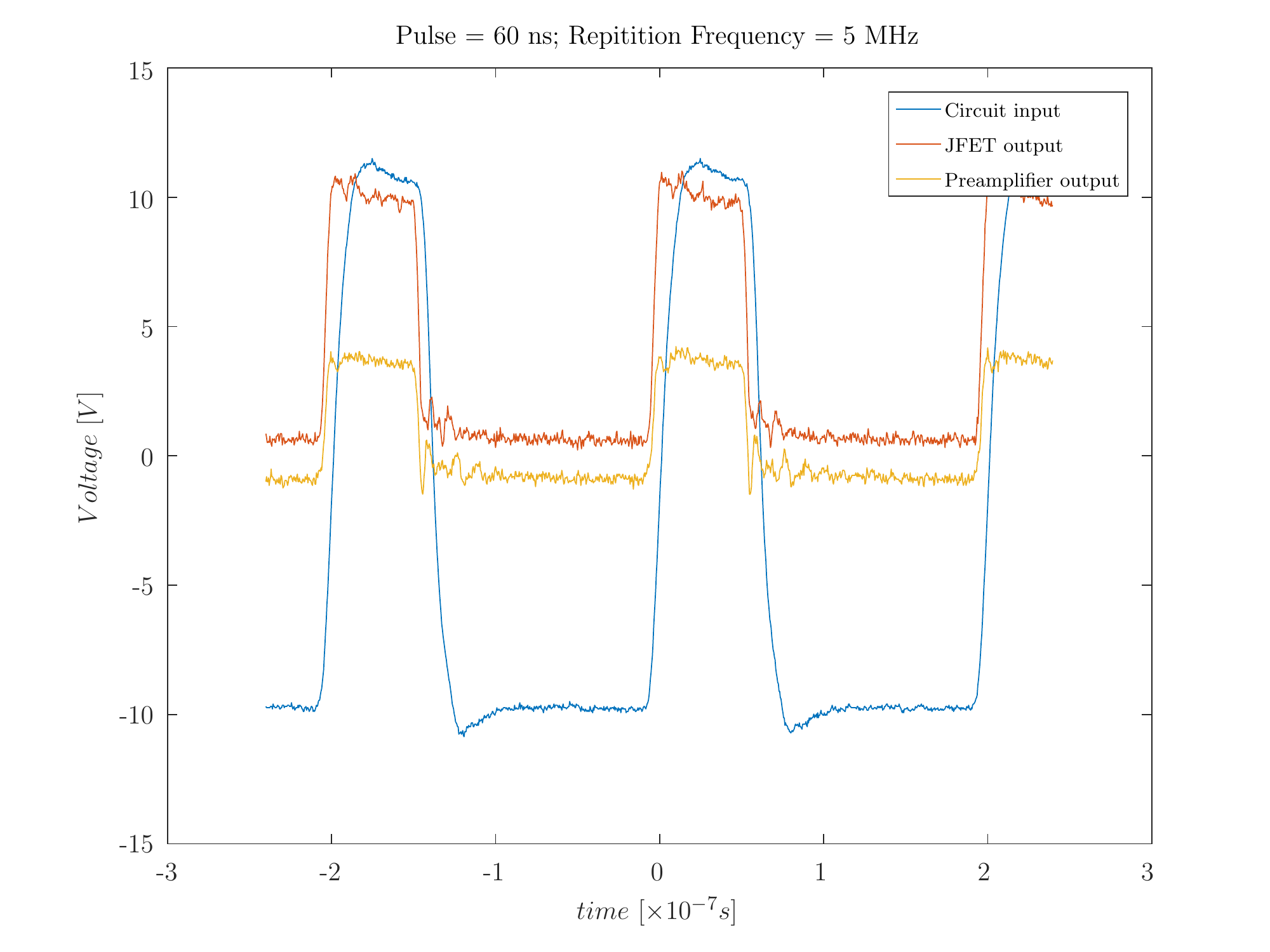}
\par\end{centering}
\centering{}\includegraphics[scale=0.7]{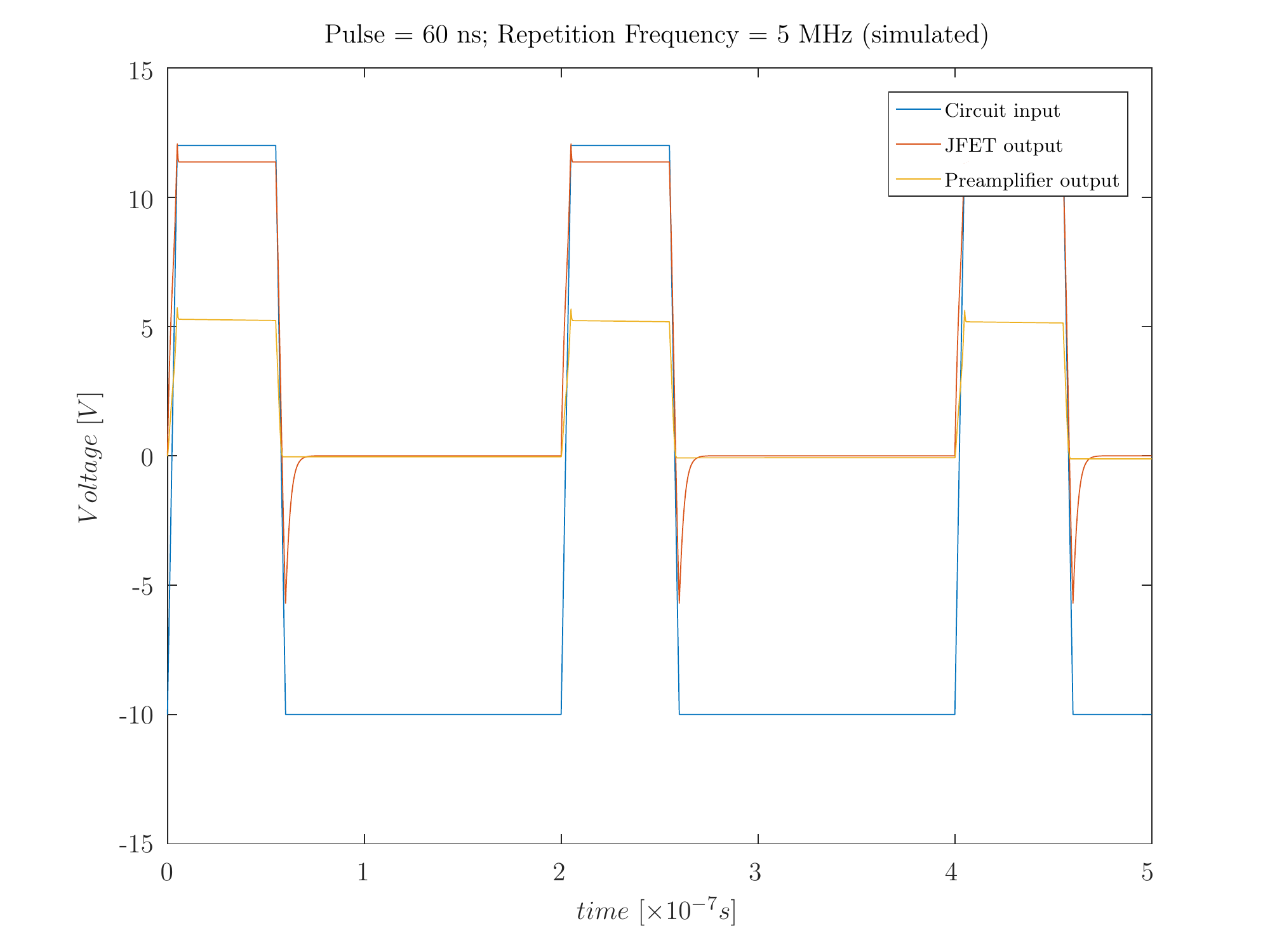}\caption{\label{fig:JFET-NMOS-60ns}JFET-NMOS transistor preamplifier response
to $60ns$ pulses}
\end{figure}

\begin{figure}[h]
\begin{centering}
\includegraphics[scale=0.7]{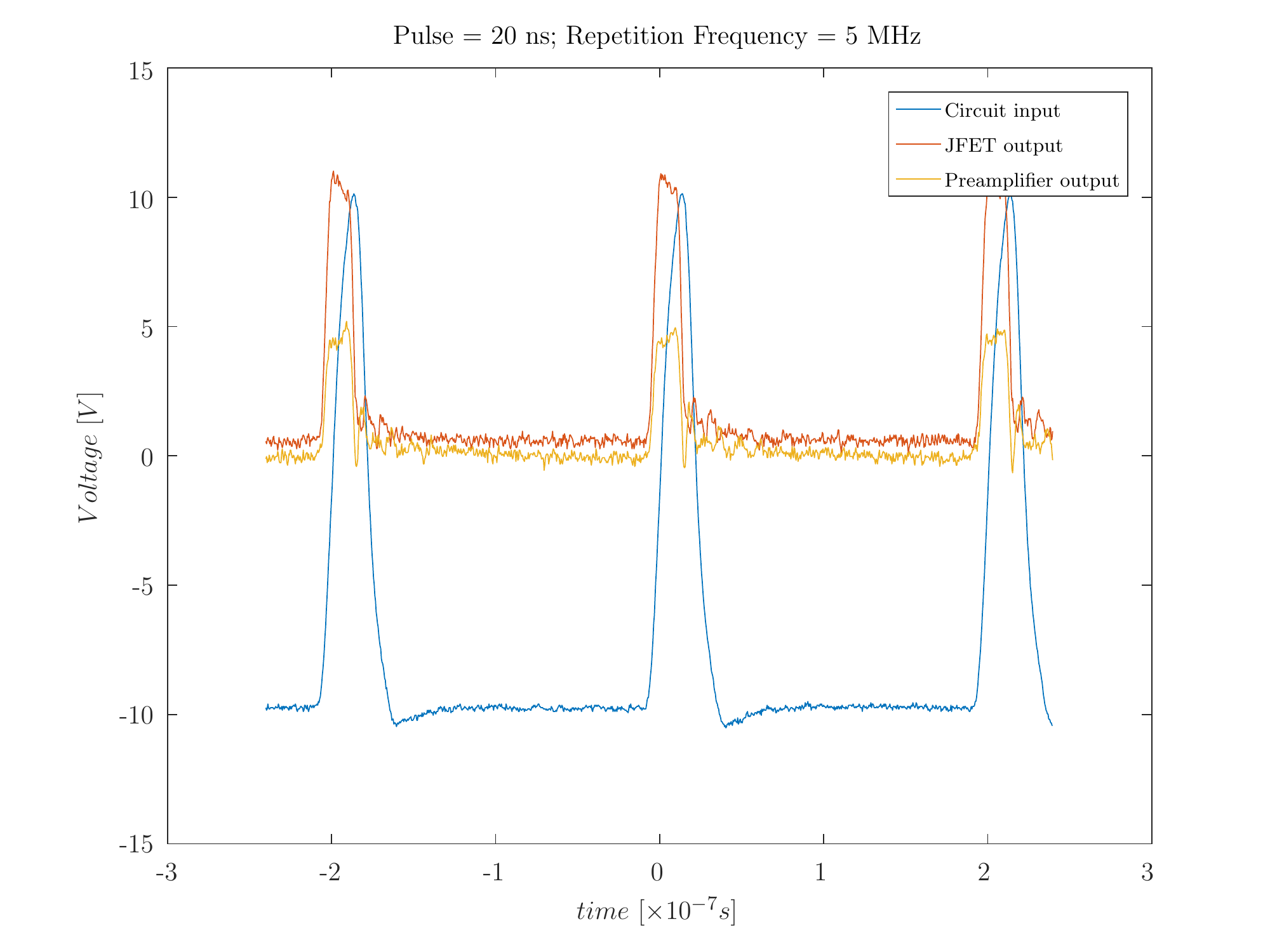}
\par\end{centering}
\centering{}\includegraphics[scale=0.7]{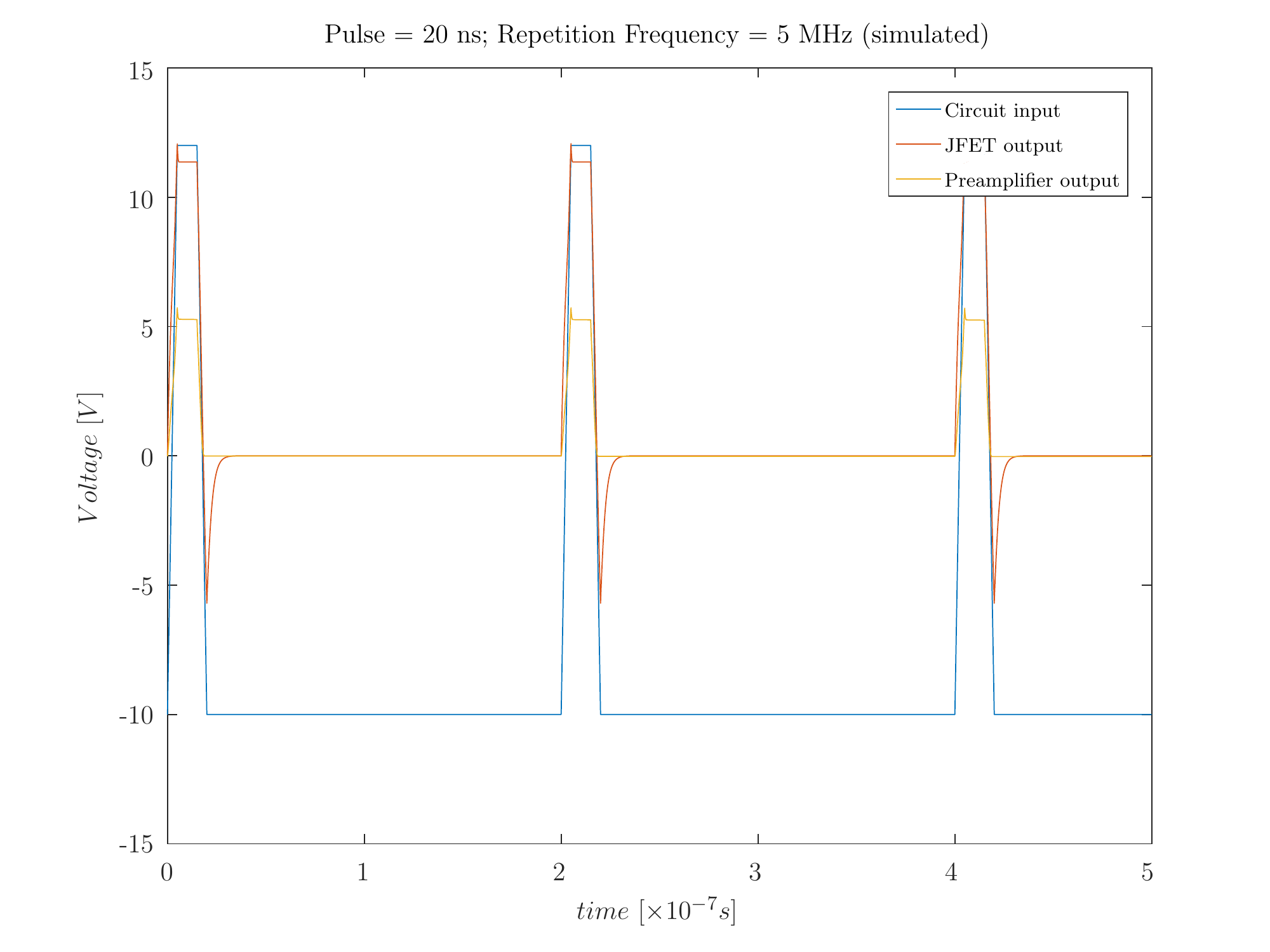}\caption{\label{fig:JFET-NMOS-preamplifier-20ns}JFET-NMOS transistor preamplifier
response to $20ns$ pulses}
\end{figure}

\begin{figure}[h]
\begin{centering}
\includegraphics[scale=0.7]{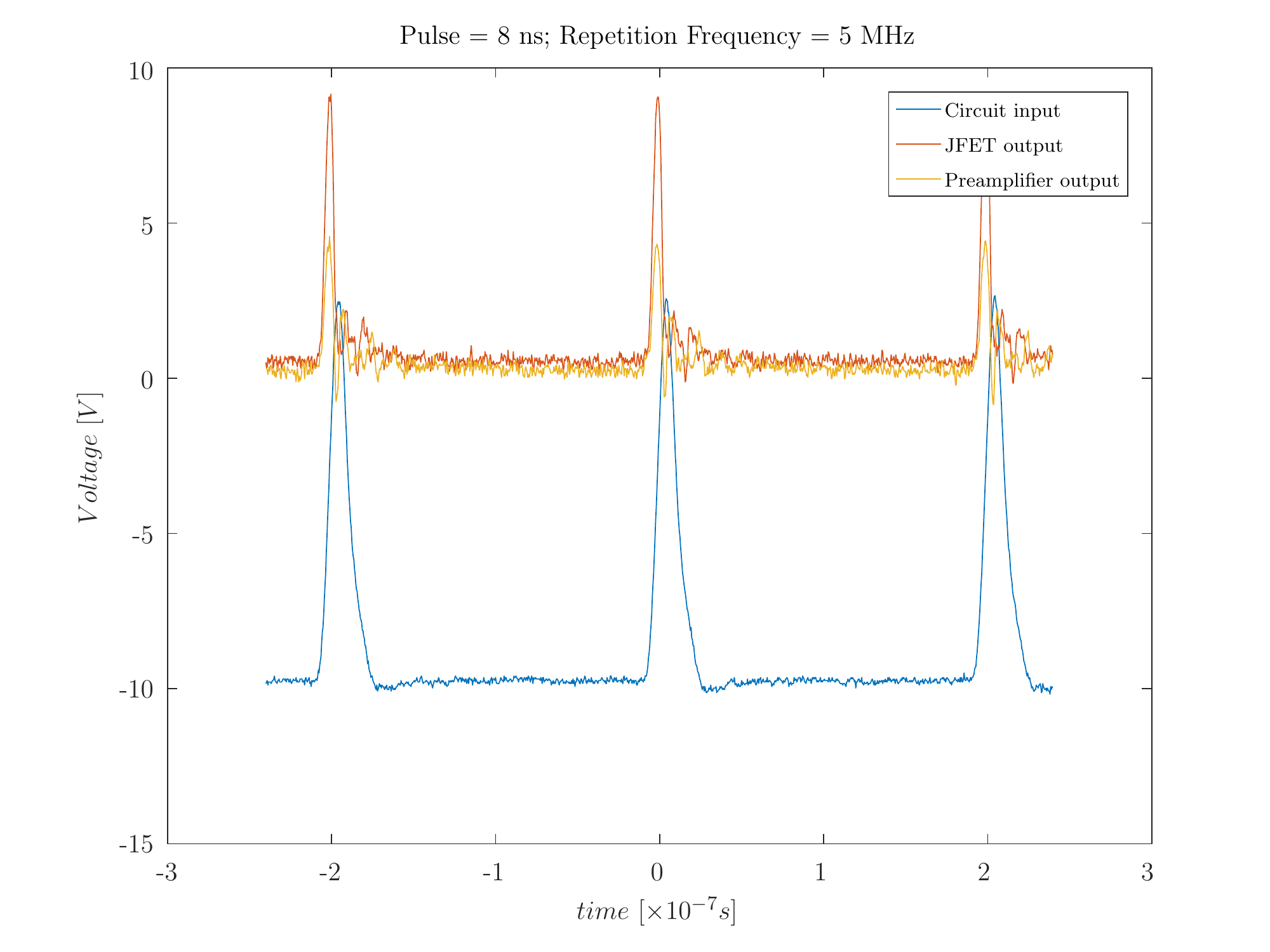}
\par\end{centering}
\begin{centering}
\includegraphics[scale=0.7]{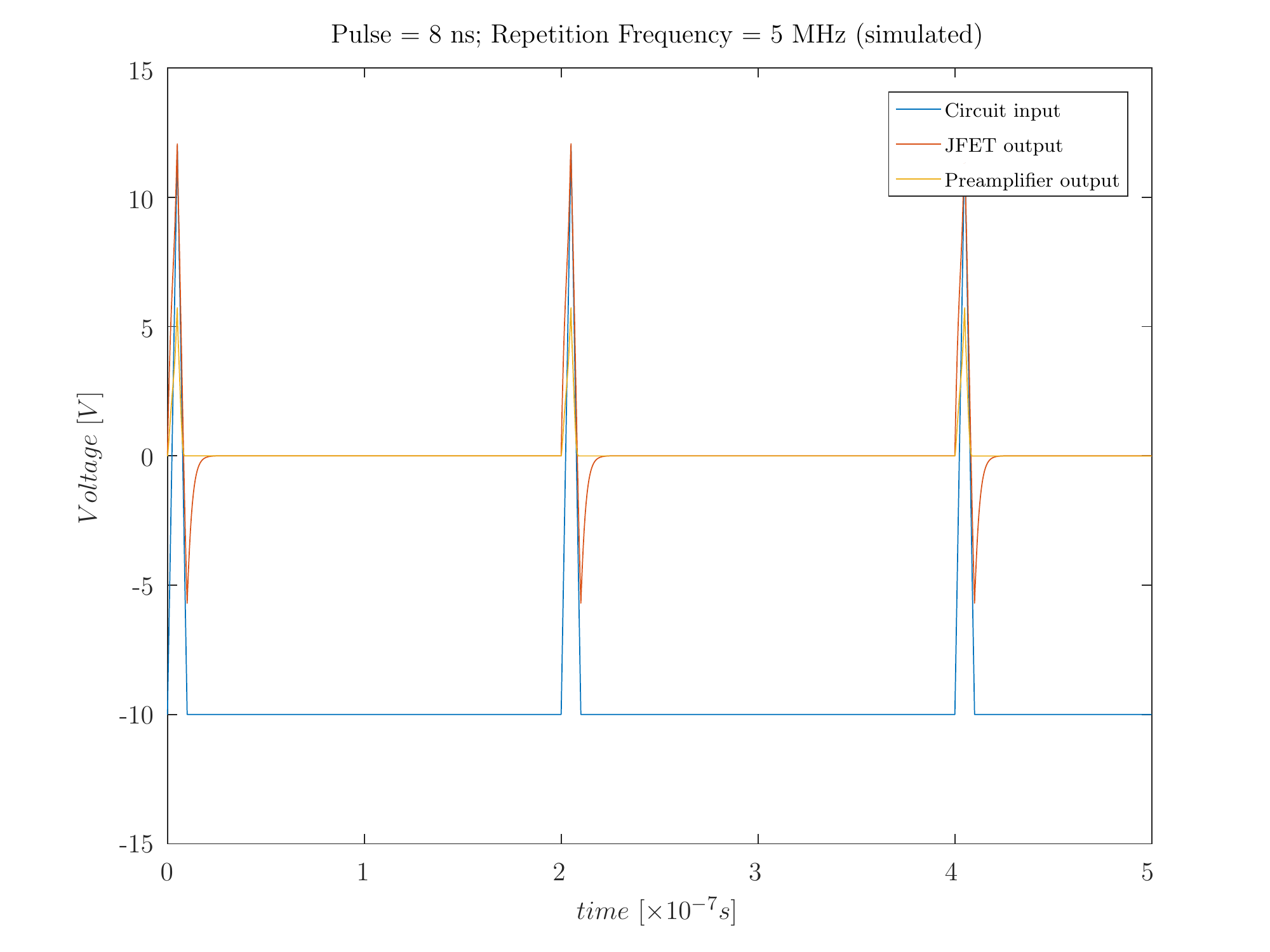}
\par\end{centering}
\centering{}\caption{\label{fig:JFET-NMOS-preamplifier-8ns}JFET-NMOS transistor preamplifier
response to $8ns$ pulses}
\end{figure}

Next we test the 2N4416 JFET at the input with a standard 2N7000 NMOS
at the output. Figures (\ref{fig:JFET-NMOS-60ns}-\ref{fig:JFET-NMOS-preamplifier-8ns})
illustrate the measured response of the preamplifier to a repetitive
$5MHz$ pulses of 60, 20 and 8 ns.

Once again the time response of the circuit in regards to the pulse
is fast, displaying a faster response than the $5\,ns$ rise time
of the generator. In this instance, the NMOS output tracks the JFET
output without a noticeable delay at any of the pulse widths. This
behavior is predicted by the simulation, deviating only on the account
that the physical JFET does not display an overshoot when switched
off. It can indicate that the actual JFET has either a much higher
resistance, or a lower capacitance than the parameters given in the
simulation.

\subsubsection{GaN HEMT single stage}

\begin{figure}[h]
\begin{centering}
\includegraphics[scale=0.7]{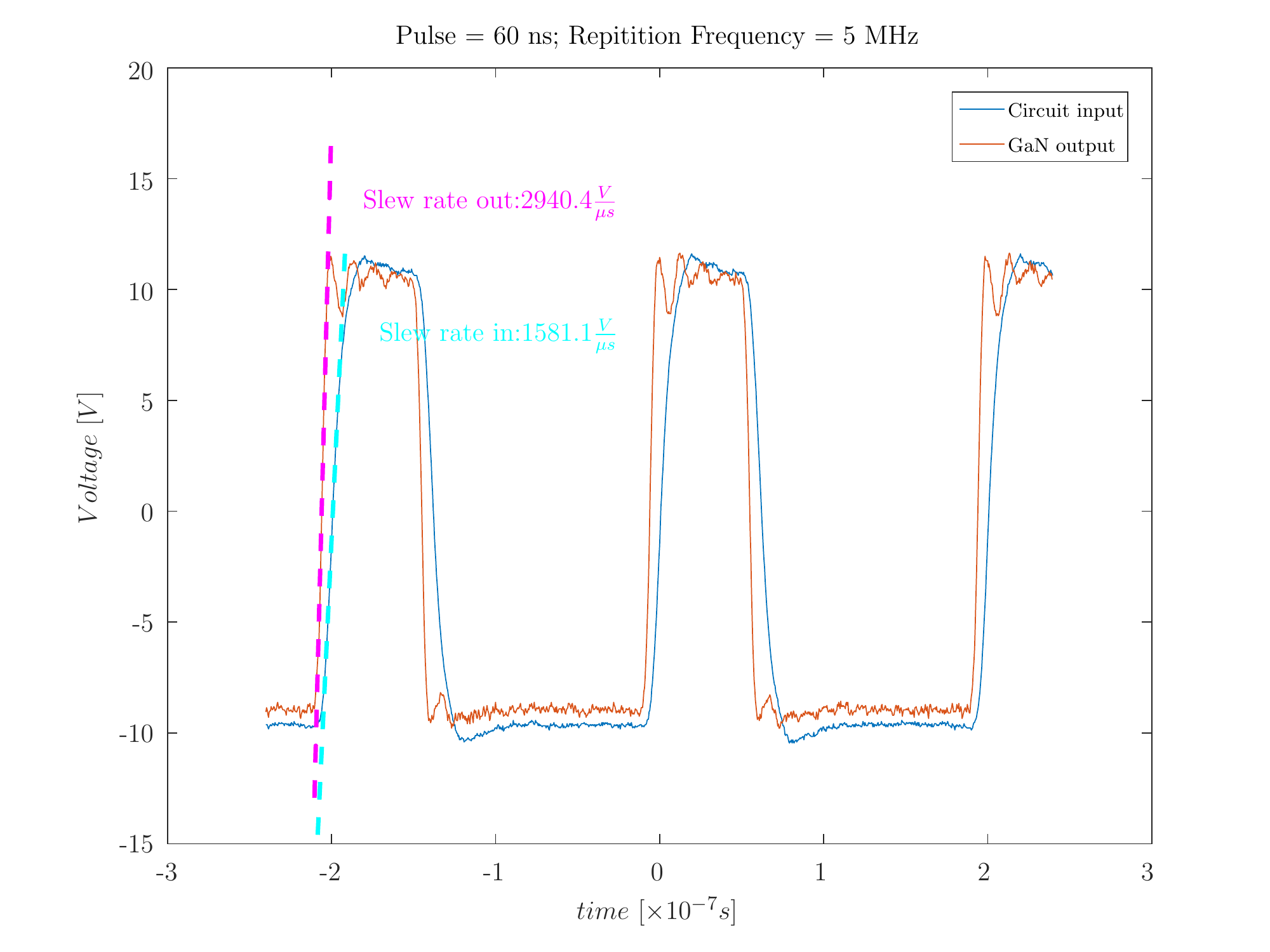}
\par\end{centering}
\centering{}\includegraphics[scale=0.7]{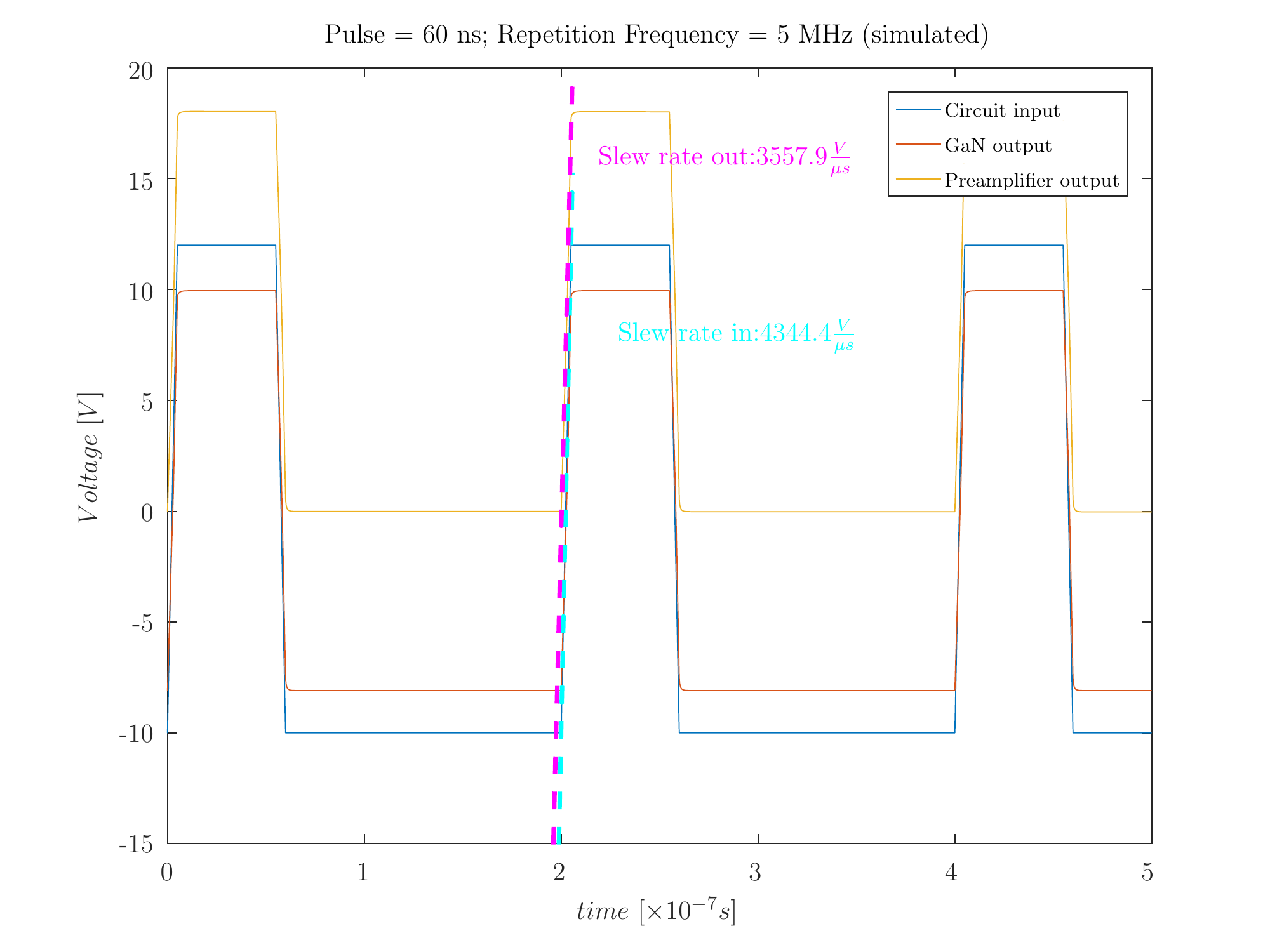}\caption{\label{fig:GaN-60ns}GaN transistor preamplifier response to $60ns$
pulses}
\end{figure}

\begin{figure}[h]
\begin{centering}
\includegraphics[scale=0.7]{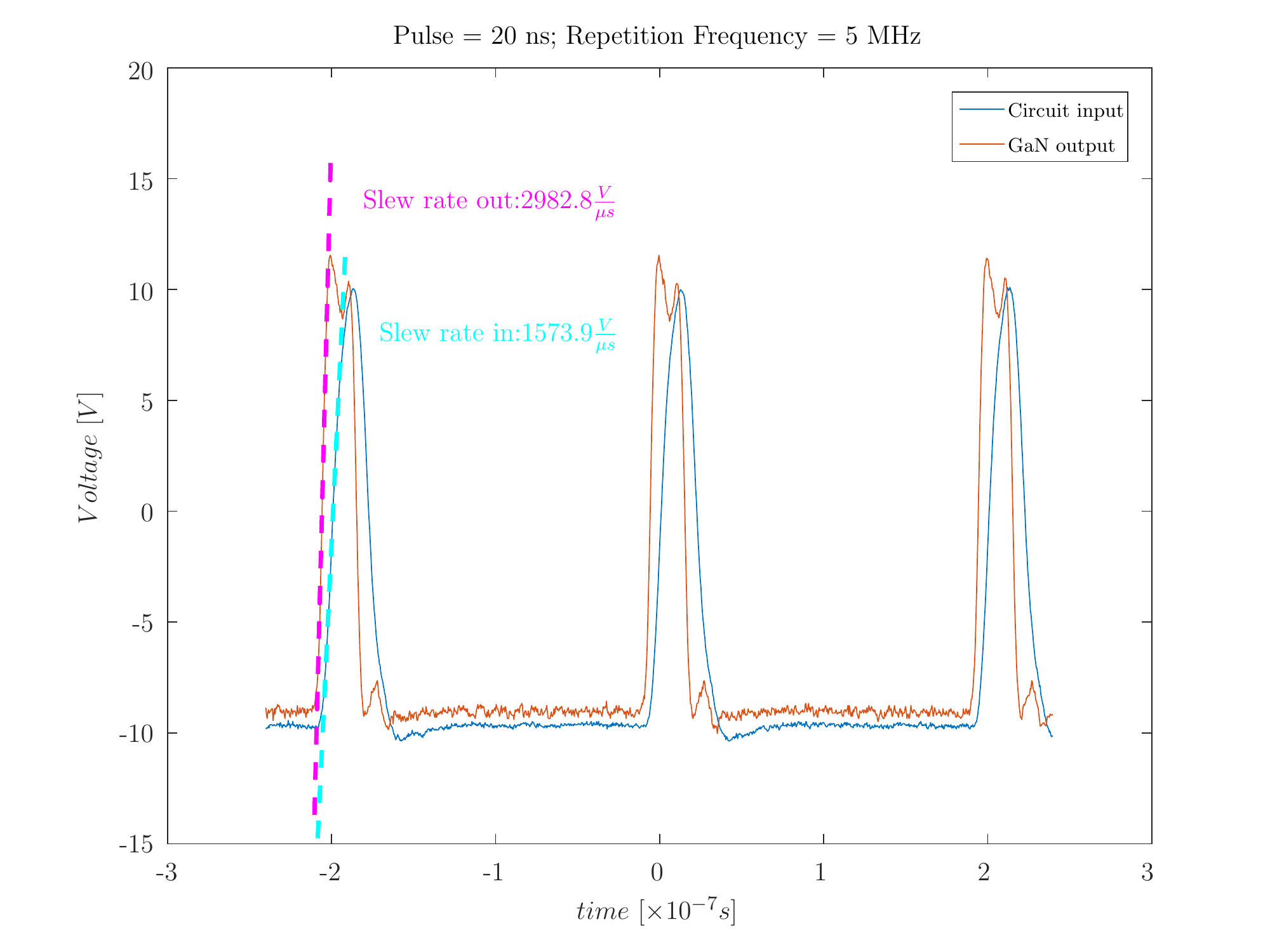}
\par\end{centering}
\centering{}\includegraphics[scale=0.7]{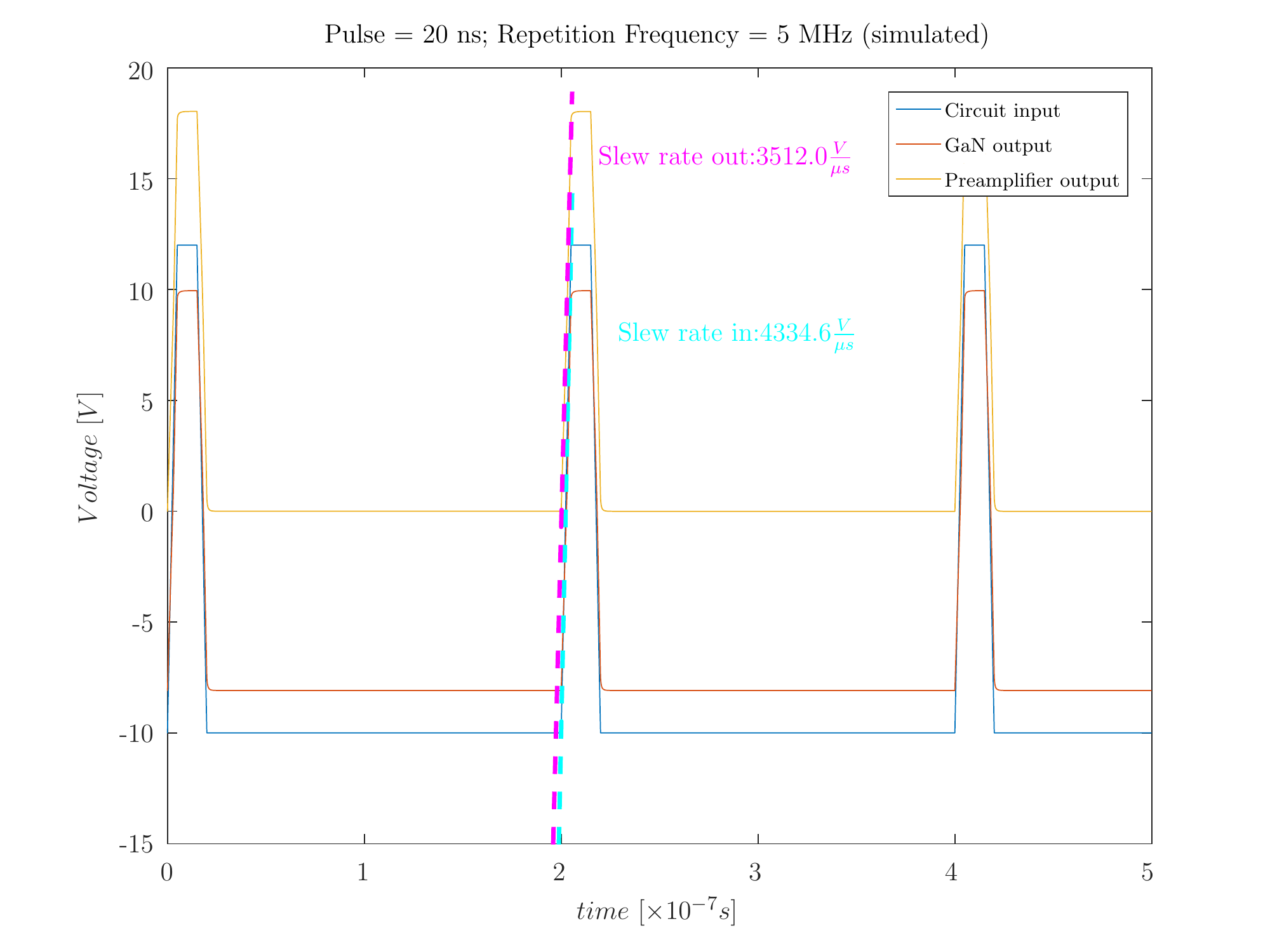}\caption{\label{fig:GaN-preamplifier-20ns}GaN transistor preamplifier response
to $20ns$ pulses}
\end{figure}

\begin{figure}[h]
\begin{centering}
\includegraphics[scale=0.7]{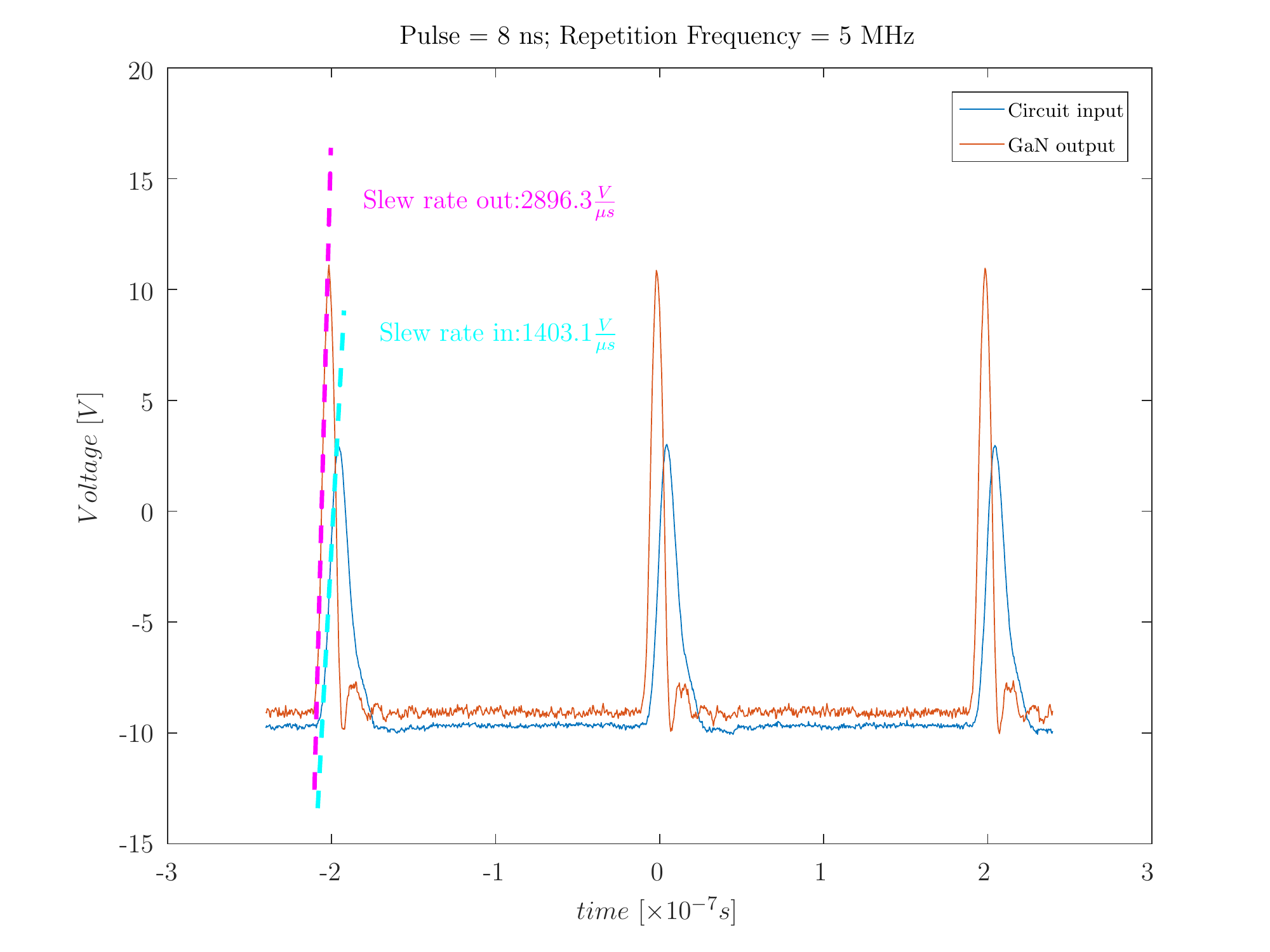}
\par\end{centering}
\begin{centering}
\includegraphics[scale=0.7]{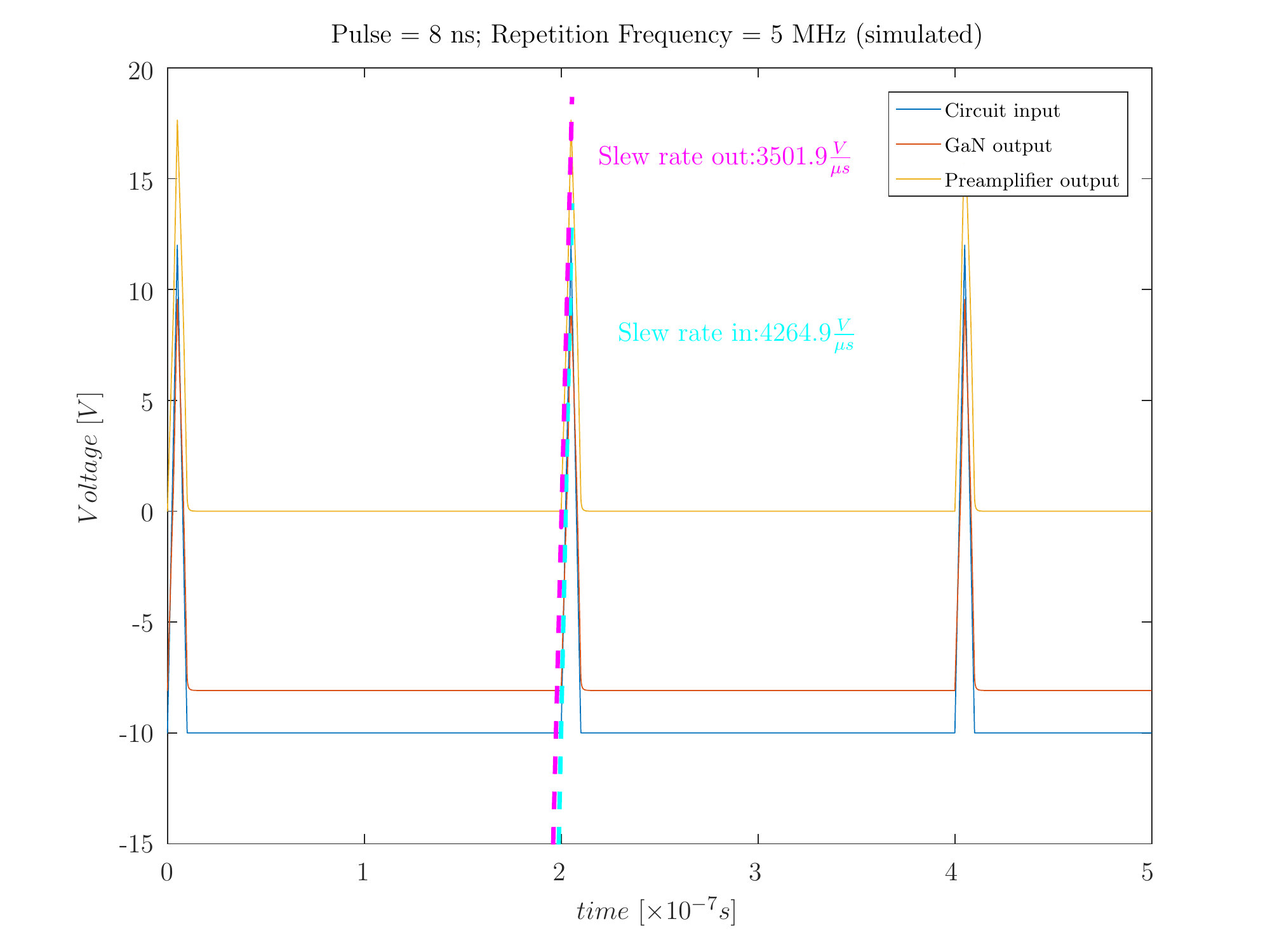}
\par\end{centering}
\centering{}\caption{\label{fig:GaN-preamplifier-8ns}GaN transistor preamplifier response
to $8ns$ pulses}
\end{figure}

As this work lays the basis for the GaN HEMT as a hardened transistor
in a preamplifier circuit, we tested it in a single stage configuration.
In many circumstances this may suffice, providing as we shall see,
a much faster response time with high sensitivity. The circuit was
realized using an EPC2038 GaN HEMT. The response of the single stage
GaN preamplifier is illustrated in figures (\ref{fig:GaN-60ns}-\ref{fig:GaN-preamplifier-8ns}).
As the sensitivity of the GaN to the input is very strong we compared
the rate of change of the voltage at the input and output, as it is
obvious from the measured results that it is much faster at the output.
We will refer to this rate of voltage change (maybe in an unorthodox
manner) as the slew rate. In order to avoid cluttering the graphs
and keeping them clear we opted to omit the output at the preamplifier
(after the capacitor) and display only the measured output at the
GaN transistor. The behavior at the output of the preamplifier, is
similar to the behavior shown in the simulation.

From the figures \ref{fig:GaN-60ns}-\ref{fig:GaN-preamplifier-8ns},
it is obvious that in the real world the GaN's slew rate is higher
than the slew rate of the generator that we used. According to the
displayed figures one can see that the GaN's slew rate at $\sim3000\,V/\mu s$
is approximately double than that of the source at $\sim1500\,V/\mu s$.
On the other hand, looking at the simulated results, while we may
say that the model approximates the GaN's behavior with a slew rate
of $\sim3500\,V/\mu s$. It is obvious that the Frequency generator
has less than half the specified slew rate of $\sim4300\,V/\mu s$,
thus in the experimental results it seems to lag behind the transistors.

\section{Discussion}

Having a Wide bandgap of $3.44\,eV$, a critical field strength of
$6\times10^{6}\,V/cm$ and high electron mobility of $2000\,\nicefrac{cm^{2}}{V\cdot s}$
places the GaN HEMPT as a very promising material for fast and hardened
transistors required for radiation detectors. Further increasing its
frequency performance, is creating a layered structure resulting in
a high electron mobility transistor (HEMT). Such transistors are now
commercially available, on both silicon and silicon carbide substrates
mostly oriented to fast switching power applications. Our application
is somewhat different as we intend to use the transistors for amplifying
small signals in an environment with high energy particles and radiation,
for which we need a radiation hardened and wide bandwidth preamplifier
responding to short events, thus increasing our time resolution. The
response of several preamplifier circuits were examined by both simulation
and practical circuits. The first was a simple two stage amplifier
with a wide bandwidth silicon JFET at the input and a bipolar transistor
at the output. This was followed by a circuit consisting of a JFET
at the input and a NMOS FET at the output. In a following simulation
the input transistor was replaced by a GaN HEMT which is commercially
available. The simulation results illustrate the superior performance
of the GaN HEMT reliably amplifying pulses without much distortion
down to pulse widths of 200 ps. As a whole we see the response of
the physical circuits agreeing well with the simulations. Both the
JFET input circuits and the GaN preamplifier display a fast response
to the input displaying a much faster slew rate compared to the input
signal. It was found that the actual slew rate of the GaN was approximately
$3000\,V/\mu s$ which is $17\%$ lower than that is predicted by
the simulation. To our best knowledge no such comparative study was
previously published.

This experimental work consists of some of the required underlying
work for fabrication of preamplifiers for reliability testing while
being exposed to $\gamma$ radiation. Following this study, we will
analyze the above circuits while being exposed to a $Co^{60}$ source. 

\bibliographystyle{plain}
\bibliography{../../References}

\begin{thebibliography}{10}

\bibitem{ayzenshtat2011measurement}
GI~Ayzenshtat, VG~Bozhkov, and A~Yu Yushchenko.
\newblock Measurement of the electron saturation velocity in an {AlGaAs/InGaAs}
  quantum well.
\newblock {\em Russian Physics Journal}, 53(9), 2011.

\bibitem{EPC2038}
Efficient~Power Conversion.
\newblock {EPC2038} - {Enhancement Mode Power Transistor}, 2021.

\bibitem{golan2018improved}
Gady Golan, Moshe Azoulay, Tsuriel Avraham, Ilan Kremenetsky, and Joseph~B
  Bernstein.
\newblock An improved reliability model for {Si} and {GaN} power {FET}.
\newblock {\em Microelectronics Reliability}, 81:77--89, 2018.

\bibitem{golan2018novel}
Gady Golan, Moshe Azoulay, Saleh Shaheen, and Joseph~B Bernstein.
\newblock A novel reliability model for {GaN} power {FET}.
\newblock In {\em 2018 IEEE International Conference on the Science of
  Electrical Engineering in Israel (ICSEE)}, pages 1--5. IEEE, 2018.

\bibitem{harris2011commercial}
Christopher Harris, Raymond Pengelly, S~Sheppard, T~Smith, B~Pribble, S~Wood,
  and C~Platis.
\newblock Commercial gan devices for switching and low noise applications.
\newblock In {\em CS MANTECH Conference}, 2011.

\bibitem{kabouche2019high}
Riad Kabouche, Romain Pecheux, Kathia Harrouche, Etienne Okada, Farid Medjdoub,
  Joff Derluyn, Stefan Degroote, Marianne Germain, Filip Gucmann, Callum
  Middleton, et~al.
\newblock High efficiency aln/gan hemts for q-band applications with an
  improved thermal dissipation.
\newblock {\em International Journal of High Speed Electronics and Systems},
  28(01n02):1940003, 2019.

\bibitem{kamp1999gan}
Markus Kamp, C~Kirchner, V~Schwegler, A~Pelzmann, Karl~Joachim Ebeling,
  M~Leszczynski, I~Grzegory, T~Suski, and S~Porowski.
\newblock {GaN} homoepitaxy for device applications.
\newblock {\em Materials Research Society Internet Journal of Nitride
  Semiconductor Research}, 4(S1):878--889, 1999.

\bibitem{kim2020effects}
Hyun-Seop Kim, Myoung-Jin Kang, Jeong~Jin Kim, Kwang-Seok Seo, and Ho-Young
  Cha.
\newblock Effects of recessed-gate structure on algan/gan-on-sic mis-hemts with
  thin aloxny mis gate.
\newblock {\em Materials}, 13(7):1538, 2020.

\bibitem{kuzuhara2016algan}
Masaaki Kuzuhara, Joel~T Asubar, and Hirokuni Tokuda.
\newblock {AlGaN/GaN} high-electron-mobility transistor technology for
  high-voltage and low-on-resistance operation.
\newblock {\em Japanese Journal of Applied Physics}, 55(7):070101, 2016.

\bibitem{li2020gan}
Lei Li, Kazuki Nomoto, Ming Pan, Wenshen Li, Austin Hickman, Jeffrey Miller,
  Kevin Lee, Zongyang Hu, Samuel~James Bader, Soo~Min Lee, et~al.
\newblock Gan hemts on si with regrown contacts and cutoff/maximum oscillation
  frequencies of 250/204 ghz.
\newblock {\em IEEE Electron Device Letters}, 41(5):689--692, 2020.

\bibitem{microsemi2014gallium}
PPG Microsemi.
\newblock Gallium nitride ({GaN}) versus silicon carbide ({SiC}) in the high
  frequency ({RF}) and power switching applications.
\newblock {\em Digi-key}, 2014.

\bibitem{pengelly2012review}
Raymond~S Pengelly, Simon~M Wood, James~W Milligan, Scott~T Sheppard, and
  William~L Pribble.
\newblock A review of {GaN} on {SiC} high electron-mobility power transistors
  and {MMICs}.
\newblock {\em IEEE Transactions on Microwave Theory and Techniques},
  60(6):1764--1783, 2012.

\bibitem{pierron1967CoefficientGaAs}
E.~D. Pierron, D.~L. Parker, and J.~B. McNeely.
\newblock Coefficient of expansion of {GaAs}, {GaP}, and {Ga(As, P)} compounds
  from $-62\degree{C}$ to $200\degree{C}$.
\newblock {\em Journal of Applied Physics}, 38(12):4669--4671, 1967.

\bibitem{ren2003wide}
Fan Ren and John~C Zolper.
\newblock {\em Wide energy bandgap electronic devices}.
\newblock World Scientific, 2003.

\bibitem{roccaforte2019overview}
Fabrizio Roccaforte, Giuseppe Greco, Patrick Fiorenza, and Ferdinando Iucolano.
\newblock An overview of normally-off {GaN}-based high electron mobility
  transistors.
\newblock {\em Materials}, 12(10):1599, 2019.

\bibitem{STORM2016121}
D.F. Storm, M.T. Hardy, D.S. Katzer, N.~Nepal, B.P. Downey, D.J. Meyer,
  Thomas~O. McConkie, Lin Zhou, and David~J. Smith.
\newblock Critical issues for homoepitaxial gan growth by molecular beam
  epitaxy on hydride vapor-phase epitaxy-grown gan substrates.
\newblock {\em Journal of Crystal Growth}, 456:121 -- 132, 2016.
\newblock Proceeding of the 9th International Workshop on Bulk Nitride
  Semiconductors.

\bibitem{wellmann2017power}
Peter~J Wellmann.
\newblock Power electronic semiconductor materials for automotive and energy
  saving applications--{SiC}, {GaN}, {$Ga_2O_3$}, and diamond.
\newblock {\em Zeitschrift f{\"u}r anorganische und allgemeine Chemie},
  643(21):1312--1322, 2017.

\bibitem{zuck2004microstructure}
A~Zuck, M~Schieber, O~Khakhan, H~Gilboa, and Z~Burshtein.
\newblock Microstructure and energy resolution of 59.6 ke{V} $^{241}{Am}$ gamma
  absorption in polycrystalline ${HgI}_2$ detectors.
\newblock {\em IEEE Transactions on Nuclear Science}, 51(3):1250--1255, 2004.

\end{thebibliography}

\end{document}